\documentclass{article} % For LaTeX2e
\usepackage{iclr2027_conference,times}

\usepackage{fontawesome5}

\usepackage{amsmath,amsfonts,bm}

\def\eqref#1{equation~\ref{#1}}
\def\1{\bm{1}}

\DeclareMathAlphabet{\mathsfit}{\encodingdefault}{\sfdefault}{m}{sl}
\SetMathAlphabet{\mathsfit}{bold}{\encodingdefault}{\sfdefault}{bx}{n}

\usepackage{hyperref}
\usepackage{url}
\usepackage{pifont}
\usepackage[utf8]{inputenc} % allow utf-8 input
\usepackage[T1]{fontenc}    % use 8-bit T1 fonts
\usepackage{hyperref}       % hyperlinks
\usepackage{url}            % simple URL typesetting
\usepackage{booktabs}       % professional-quality tables
\usepackage{amsfonts}       % blackboard math symbols
\usepackage{nicefrac}       % compact symbols for 1/2, etc.
\usepackage{microtype}      % microtypography
\usepackage{comment}
\usepackage{graphicx} % Required for inserting images
\usepackage{algorithm}
\usepackage{listings}
\usepackage{amsmath, amssymb}

\usepackage{xcolor}

\definecolor{codecomment}{rgb}{0.25, 0.50, 0.75}  % light blue / steel-blue
\definecolor{codekeyword}{rgb}{0.78, 0.10, 0.18}  % red
\definecolor{codestring}{rgb}{0.10, 0.45, 0.10}   % green (unused but defined)
 
\lstdefinestyle{pseudocode}{
    backgroundcolor   = \color{white},
    basicstyle        = \ttfamily\scriptsize,
    keywordstyle      = \color{codekeyword}\bfseries,
    commentstyle      = \color{codestring},
    stringstyle       = \color{codestring},
    showstringspaces  = false,
    keepspaces        = true,
    columns           = fullflexible,
    language          = Python,
    morekeywords      = {def, return, for, in, with, as, not, None,
                         True, False, and, or, if, else, elif},
    mathescape        = false,
    breaklines        = false,
    aboveskip         = 0pt,
    belowskip         = 0pt,
}

\usepackage{caption} 
\usepackage{subcaption}
\usepackage{makecell}
\usepackage{wrapfig}

\usepackage{mathtools}

\usepackage{amsmath}

\usepackage{colortbl}
\usepackage{adjustbox}
\usepackage{arydshln}
\usepackage{multirow}

\usepackage{xcolor}
\usepackage[most]{tcolorbox}

\newtcolorbox{boxaccent}[2][]{
  enhanced, breakable,
  colback=nepapanel, colframe=nepapanel,
  boxrule=0pt, arc=0pt, outer arc=0pt,
  borderline west={3.5mm}{0pt}{nepapink},
  title=#2,
  coltitle=nepadark, colbacktitle=nepapanel,
  fonttitle=\bfseries\large\sffamily,
  titlerule=0pt,
  boxsep=0pt,
  attach boxed title to top left={xshift=6mm, yshift=-3.5mm},
  boxed title style={empty, boxrule=0pt},
  drop shadow={nepadark!30},
  top=4mm, bottom=2mm, left=6mm, right=4mm,
  #1
}

\definecolor{nepapink}{HTML}{F72585}
\definecolor{nepadark}{HTML}{333333}
\definecolor{nepapanel}{HTML}{F6F8FB}
\definecolor{nepadiagonal}{HTML}{3A86FF}

\definecolor{ftcolor}{rgb}{1,1,1} % almond
\definecolor{baselinecolor}{rgb}{1,1,1}
\definecolor{contrcolor}{rgb}{1.0, 0.98, 0.8}
\definecolor{predcolor}{rgb}{0.74, 0.83, 0.9}
\definecolor{decorrcolor}{rgb}{0.98, 0.85, 0.87} % palepink
\definecolor{knowcolor}{rgb}{0.80, 0.94, 0.75}
\definecolor{offlinecolor}{rgb}{1,1,1}
\definecolor{firsttaskcolor}{rgb}{0.97, 0.97, 0.97}
\definecolor{azure(colorwheel)}{rgb}{0.0, 0.5, 1.0}
\definecolor{gray(x11gray)}{rgb}{0.75, 0.75, 0.75}
\definecolor{lightgray}{rgb}{0.90, 0.90, 0.90}
\definecolor{darkgray}{RGB}{128,128,128}
\definecolor{teagreen}{rgb}{0.82, 0.94, 0.75}
\definecolor{almondlow}{RGB}{252,239,219} 
\definecolor{almondmiddle}{RGB}{237,225,206} 
\definecolor{almondhigh}{RGB}{224,213,194} 
\definecolor{almondultra}{RGB}{214,204,186} 
\definecolor{greylow}{RGB}{235,235,235} 
\definecolor{greymiddle}{RGB}{211,211,211} 
\definecolor{greyhigh}{RGB}{192,192,192} 
\definecolor{pinksecondbest}{RGB}{252, 241, 241}
\definecolor{greenpatch}{RGB}{0,204,0} 

\definecolor{pastelred}{RGB}{232, 131, 131}
\definecolor{pastelviolet}{RGB}{234, 229, 246}
\definecolor{champagne}{rgb}{0.97, 0.91, 0.81}
\definecolor{lightblue}{RGB}{225, 241, 255}

\definecolor{violetlight}{RGB}{225,213,231}
\definecolor{violetmiddle}{RGB}{200,174,213}
\definecolor{yellow}{RGB}{253,253,216}

\definecolor{green_ablation}{RGB}{230,255,230}
\definecolor{orange_ablation}{RGB}{255,233,215}
\definecolor{blue_ablation}{RGB}{226,247,254}
\definecolor{pink_ablation}{RGB}{254,227,239}
\definecolor{yellow_ablation}{RGB}{255,253,236}
\definecolor{violetblue_ablation}{RGB}{230,230,255}
\definecolor{grey_ablation}{RGB}{230,230,230}

\usepackage{soulutf8} 
\newcommand{\hlc}[2]{{\sethlcolor{#1}\hl{#2}}}

\title{Listening Forward: Next Patch Embedding Prediction Enables Scalable Audio Learners}

\author{%
\begin{minipage}[t]{\dimexpr\textwidth-2\tabcolsep\relax}
\centering\normalfont
{\bfseries Umberto Cappellazzo\textsuperscript{1}, Xubo Liu\textsuperscript{2}, 
Stavros Petridis\textsuperscript{1}, Maja Pantic\textsuperscript{1}} \\[0.6em]
\textsuperscript{1}Imperial College London \quad 
\textsuperscript{2}University of Surrey \\[0.4em]
\texttt{u.cappellazzo@imperial.ac.uk} \\[0.4em]
\faGlobe~\href{https://umbertocappellazzo.github.io/nape}{Project Page} \quad
\faGithub~\href{https://github.com/umbertocappellazzo/nape/tree/main}{Code}
\end{minipage}%
}

\iclrfinalcopy % Uncomment for camera-ready version, but NOT for submission.
\begin{document}

\maketitle

\begin{abstract}
Self-supervised learning (SSL) has driven substantial progress in audio representation learning, though existing methods have increasingly relied on elaborate pre-training recipes to reach competitive performance. A markedly different pre-training philosophy underpins the most influential progress in language modeling and, more recently, in visual representation learning: rather than train encoders as static feature extractors, models are trained to predict the next element, a discrete token or a continuous embedding, from the preceding context. Autoregressive prediction thereby provides a unified pre-training interface that transfers across modalities, compelling the model to learn the underlying data distribution. We ask whether such a simple causal paradigm can yield strong audio learners, given that audio's temporal structure makes autoregressive prediction of patch embeddings a natural fit. We introduce NAPE (\textbf{N}ext-\textbf{A}udio-\textbf{P}atch-\textbf{E}mbedding prediction), a self-supervised framework in which a causal Transformer predicts each next patch embedding of a log-mel spectrogram from the previous ones, using causal masking and stop-gradient as its sole training signal. The design is intentionally minimalist, avoiding reconstruction decoders, acoustic tokenizers, student-teacher setups, and auxiliary regularization losses. Across six audio and speech benchmarks, NAPE achieves state-of-the-art fine-tuning performance on several tasks, scales consistently across encoder sizes, and yields strong linear-probing results. NAPE also produces structured attention patterns without explicit supervision.
\end{abstract}

\section{Introduction}
\label{sec:introduction}

Self-supervised learning (SSL) has emerged as a foundational
paradigm for representation learning across modalities, delivering strong transfer performance without the cost of human annotations~\citep{radford2018improving, devlin2019bert, brown2020language, chen2021simsiam, he2022masked, oquab2023dinov2, baevski2020wav2vec, kong2024audio, chen2024eat}. In audio, SSL methods largely adapt
paradigms first developed in vision. Vision Transformer
backbones~\citep{dosovitskiy2020image} are pre-trained on
spectrograms either through \textit{masked-spectrogram
modeling}~\citep{niizumi2022masked, huang2022masked, chong2023masked, chen2022beats, dinkel2024scaling, niizumi2026rethinking} or through \textit{student-teacher distillation} objectives~\citep{ahmed2024asit, chen2024eat, alex2025sslam, yang2025spear}, with domain-specific augmentations tailored to the time-frequency structure of audio. Most of these methods pre-train on AudioSet~\citep{gemmeke2017audio} and are evaluated primarily by fine-tuning performance on downstream classification tasks. 

These paradigms, however, rely on increasingly elaborate pre-training recipes. Existing methods typically depend on reconstruction decoders that map latent features back to raw mel content~\citep{huang2022masked, chen2024eat}, separately-trained acoustic tokenizers that supply discrete semantic
targets~\citep{chen2022beats}, teacher-student setups with
exponential moving average (EMA)-updated encoders~\citep{chen2024eat, alex2025sslam, ahmed2024asit}, auxiliary regularization losses that stabilize training~\citep{fei2023jepa, alex2025sslam}, and multi-codebook vector quantisation~\citep{yang2025spear}. Recently, a distinct line of work has begun to reshape SSL by shifting away from reconstruction toward the direct prediction of latent embeddings. This shift extends a paradigm long established in language modeling, where models are trained not as static feature extractors, but as predictive systems that model the data distribution through a single causal objective. Autoregressive prediction has thereby provided a unified pre-training interface across modalities, from discrete tokens in language to continuous embeddings in vision. Two variants of this philosophy have gained traction. \emph{Joint-embedding predictive} approaches~\citep{assran2023self, oquab2023dinov2, fei2023jepa, bardes2024revisiting, balestriero2025lejepa, yuksel2025wavjepa, huang2026text, wu2026visreg} predict the latent embeddings of masked regions from a context view, produced by a target branch that is typically maintained via an EMA of the online encoder or stabilized by auxiliary regularization losses. While these methods are scalable and achieve strong performance, most of them rely on heavy heuristics (e.g., EMA, frozen layers, teacher-student architectures) to ensure training stability. \emph{Autoregressive next-embedding} approaches~\citep{teoh2025next, xu2025nepa, bredis2026next, yao2026rethinking, maes2026leworldmodel}, in contrast, predict each latent embedding directly from the preceding ones, mirroring the causal next-token objective that drives modern large language models~\citep{radford2018improving, brown2020language, liu2024deepseek, yang2025qwen3}. Both variants have quickly become a promising direction for representation learning in \textit{vision}.

Despite this progress, predictive next-embedding methods \textit{remain absent} from audio SSL, and joint-embedding predictive approaches themselves have seen \textit{only limited exploration} in the audio and speech settings~\citep{fei2023jepa, yuksel2025wavjepa, tuncay2025audio}. The absence of any next-embedding autoregressive method for audio is particularly striking because audio is arguably the modality \emph{most} naturally suited to this paradigm. Unlike images, whose 2D structure is approximately isotropic and admits no canonical ordering, audio signals unfold along a well-defined temporal axis and sequential structure is intrinsic to the signal, not imposed on it. Predicting the next patch of a spectrogram from the past ones mirrors both how acoustic events emerge in time and how modern language models learn from sequential data. Motivated by this natural alignment, we seek a next-embedding prediction framework for audio that is deliberately minimalist while still delivering state-of-the-art downstream performance across audio and speech benchmarks.

We therefore introduce NAPE (\textbf{N}ext-\textbf{A}udio-\textbf{P}atch-\textbf{E}mbedding prediction), a self-supervised framework that brings the causal next-embedding prediction paradigm to audio. Given a log-mel spectrogram, NAPE first applies a patch embedding layer that splits the spectrogram into non-overlapping patches and projects each of them into a $d$-dimensional embedding, producing a 2D grid of patch embeddings. Since the causal Transformer that follows operates on a 1D sequence, this grid must be traversed under a \emph{scanning order}. This choice is a design axis specific to audio: unlike static images, spectrograms have a strong temporal axis and a qualitatively different frequency axis, so the order in which the grid is linearized determines both the causal context available at each prediction step and the structural inductive bias of the model. We consider four scanning strategies, depicted in Figure~\ref{fig:patch_type}. A causal Transformer encoder then processes the linearized sequence, and a lightweight predictor head produces an estimate of the next patch embedding, analogous to next-token prediction in language modeling, but operating in continuous embedding space rather than over a discrete vocabulary. Prediction quality is measured by negative cosine similarity to the target patch embedding under a stop-gradient. This objective requires no reconstruction decoder, no acoustic tokenizer, no student-teacher setup, and no auxiliary regularization losses: the entire learning signal comes from the model's ability to anticipate the next embedding in the sequence.

Extensive experiments on standard audio and speech benchmarks, including 
AudioSet~\citep{gemmeke2017audio},
ESC-50~\citep{piczak2015esc}, Speech Commands V1 and
V2~\citep{warden2018speech}, and IEMOCAP~\citep{busso2008iemocap}, show that NAPE achieves state-of-the-art performance on several tasks. NAPE exhibits favorable scaling properties across three encoder sizes: \textit{Small}, \textit{Base}, and \textit{Large}, with 19/85/303 million parameters, respectively. Beyond fine-tuning, NAPE delivers strong linear-probing results despite being a purely predictive model whose objective is not aligned with linear separability, indicating that the learned representations remain discriminative even under strict feature-freeze evaluation. A qualitative analysis of NAPE's attention patterns and embedding-space structure further confirms that NAPE learns meaningful, structured features from audio. Our main contributions are:

\begin{itemize}
    \item We introduce NAPE, the first self-supervised audio
    framework built around causal next-patch-embedding prediction, offering a substantially simpler alternative to reconstruction-based, masked-modeling, and joint-embedding predictive approaches for audio.

    \item Through systematic ablations, we identify the design
    axes that make next-embedding prediction work in the audio
    setting---scanning order, predictor head, prediction target, patch embedding layer, and the three key components of causality, prediction shift, and stop-gradient---and converge on an optimal configuration for the framework.

    \item Across six audio and speech benchmarks, NAPE achieves
    strong downstream results at three encoder scales, exhibits favorable scaling behavior, and delivers competitive linear-probing performance.

    \item We provide qualitative evidence that NAPE learns
    structured representations, with attention patterns that
    reason jointly about the current spectral context and the
    frequency-consistent temporal history, and embedding-space
    behavior that groups acoustically-similar patches without any explicit labels.
\end{itemize}

\section{The NAPE Framework}
\label{sec:methodology}

\begin{figure}[t]
    \centering
    \includegraphics[width=1\textwidth]{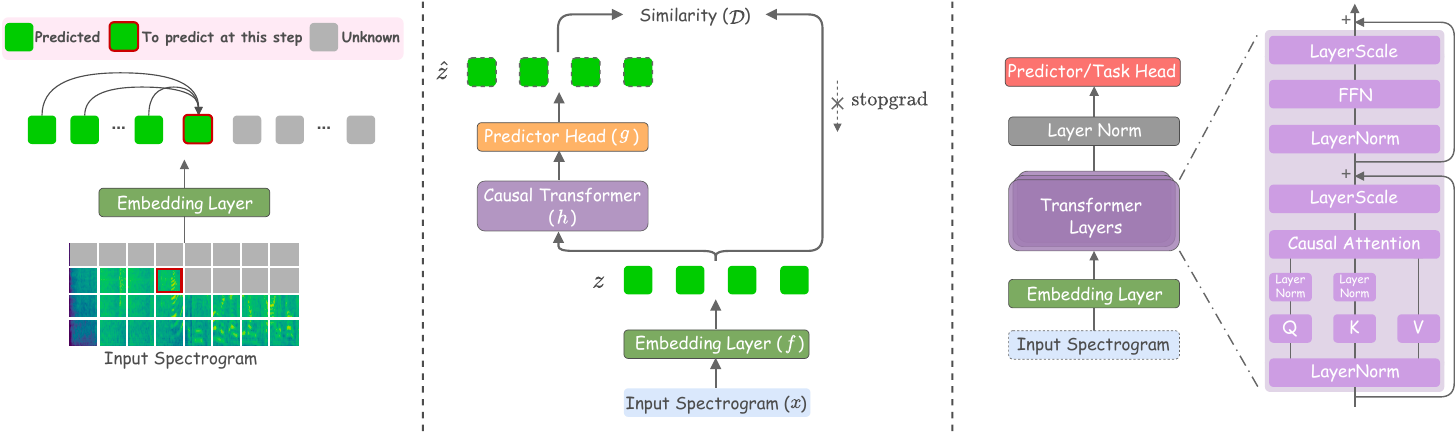}
    \caption{\textbf{Left:} NAPE's overview. The input spectrogram is split into patches and embedded into a sequence of embeddings. At each step, the model predicts the embedding of the \textcolor{red}{next patch} (red border) using only the embeddings of the \textcolor{greenpatch}{preceding patches}; patches at \textcolor{darkgray}{future positions} are hidden by the causal attention mask. \textbf{Middle:} The NAPE pipeline: patch embeddings $z$ are processed by the causal encoder $h$ and predictor $g$ to produce predictions $\hat{z}$, which are compared against the targets $z$ under stop-gradient using a similarity function $\mathcal{D}$ (i.e., negative cosine similarity). \textbf{Right:} Encoder architecture ($h$): multiple stacked Transformer layers with pre-norm design, causal self-attention, LayerScale, and query-key normalization.}
    \label{fig:architecture}
    %\vspace{-0.3cm}
\end{figure}

In this section, we describe the NAPE framework in detail. NAPE is a self-supervised pre-training method that trains a causal Transformer to predict the embedding of the next patch of a log-mel spectrogram from the preceding ones, relying only on causal masking and stop-gradient. We describe NAPE's main components and downstream adaptation in the next subsections. Figure~\ref{fig:architecture} depicts the overall framework.

\subsection{Spectrogram Patchification and Scanning Order}
\label{sec:nape-patches}

\paragraph{Input Representation.} Given a raw waveform, we compute a log-mel spectrogram $x \in \mathbb{R}^{1 \times F \times T_{\text{frames}}}$ with $F$ frequency bins and $T_{\text{frames}}$ time frames. We split $x$ into non-overlapping square patches of size $P \times P$, yielding a 2D grid of $ N = T_F \cdot T_T$ patches with $T_F = F/P$ and $T_T = T_{\text{frames}}/P$. Each patch is projected to a $d$-dimensional embedding by a patch embedding layer $f$, producing the sequence $\{z_1, \ldots, z_N\} \in \mathbb{R}^{N \times d}$.  We use a standard Conv2d patch embedding layer by default, but we also consider a deeper convolutional stem with batch normalization (\emph{convstem})~\citep{xiao2021early} and a speech-oriented stem that treats the mel axis as feature channels and applies temporal-only convolutions (\emph{speechstem})~\citep{team2026gemma}. We refer to Section~\ref{sec:ablations} for details and results about the choice of $f$.

\begin{figure}[t]
    \centering
    \includegraphics[width=1\textwidth]{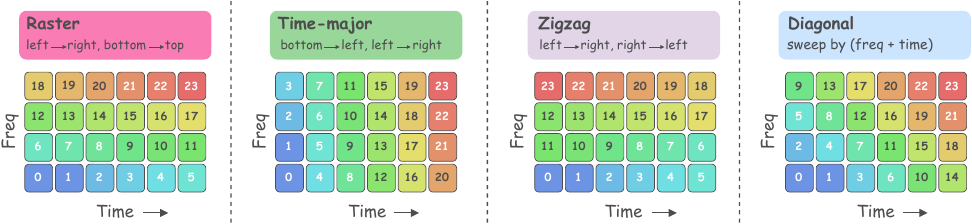}
    \caption{\textbf{Patch scanning orders.} NAPE linearizes the 2D spectrogram patch grid into a 1D causal sequence in one of four ways: \emph{raster}
    (left-to-right, bottom-to-top), \emph{time-major} (bottom-to-top within each time column, then advance in time), \emph{zigzag} (raster with alternating
    row directions), and \emph{diagonal} (sweep by frequency-plus-time index). The numbers indicate the position of each patch in the resulting sequence.}
    \label{fig:patch_type}
    %\vspace{-0.3cm}
\end{figure}

\paragraph{Scanning Order.} Transformer encoders operate on 1D sequences, so a 2D patch grid must be linearized before it can be processed. Under bidirectional attention, as used in prior audio SSL methods~\citep{gong2021ast, huang2022masked, chen2024eat}, the choice of linearization is not a functional design choice: self-attention is permutation-equivariant given positional embeddings, so any consistent ordering yields the same representations. NAPE's causal formulation, however, breaks this equivariance as the ordering determines which patches are ``past'' (visible to a given position) and which are ``future'' (to be predicted), and therefore imposes a substantive inductive bias on the model. Since spectrograms have a strong temporal axis and a qualitatively different frequency axis, unlike natural images, whose 2D structure is approximately isotropic,  the choice of scanning order for causal prediction is particularly consequential. Thus, we consider four orderings, illustrated in Figure~\ref{fig:patch_type}. \textbf{\emph{Raster}} (left-to-right, bottom-to-top) sweeps time before advancing in frequency and is the standard patch ordering used in vision and audio Transformer models, where each prediction is conditioned on the entire past time axis at every frequency traversed so far. \textbf{\emph{Time-major}} (bottom-to-top within each time column, then advance in time) sweeps frequency before advancing in time, so each prediction is conditioned on the full frequency profile of every past time step. \textbf{\emph{Zigzag}} is an alternating-direction variant of raster: rows alternate left-to-right and right-to-left, so consecutive patches in the sequence remain spatially adjacent even across row transitions. This avoids the spatial discontinuity raster incurs when jumping from the end of one row to the start of the next. \textbf{\emph{Diagonal}} sweeps patches along anti-diagonals of the grid, mixing time and frequency progression at every step and thereby encoding 2D spatial priors more uniformly than the other three orders.

\subsection{Next-Audio-Patch-Embedding Prediction}
\label{sec:nape-prediction}

\paragraph{Prediction Task.} Given the embedding sequence
$z = \{z_1, \ldots, z_N\}$ produced by $f$ in the chosen scanning order, NAPE jointly trains an encoder $h$ and a lightweight predictor head $g$ so that, at each position $t$, the model produces an estimate of the next patch embedding
$z_{t+1}$ using only the previous patches:
\begin{equation}
  \hat{z}_{t+1} = g\!\left(h(z_{\leq t})\right),
  \label{eq:nape-prediction}
\end{equation}
where $z_{\leq t} = \{z_1, \ldots, z_t\}$ denotes the patch embeddings at all positions up to and including $t$. The restriction to the past patches is enforced by a \textit{causal attention mask} that prevents each position from attending to patches at later positions in the sequence. This is directly analogous to next-token prediction in language modeling, but operates in a continuous embedding space rather than over a discrete vocabulary.

\paragraph{Loss Function.} Following SimSiam~\citep{chen2021simsiam}, we measure prediction quality by the negative cosine similarity between the target embedding $z_{t+1}$ and the predicted embedding $\hat{z}_{t+1}$:
\begin{equation}
  \mathcal{D}(z_{t+1}, \hat{z}_{t+1})
  \;=\; -\, \frac{z_{t+1}}{\lVert z_{t+1} \rVert_2}
  \cdot
  \frac{\hat{z}_{t+1}}{\lVert \hat{z}_{t+1} \rVert_2},
  \label{eq:cosine-similarity}
\end{equation}
where $\lVert \cdot \rVert_2$ is the $\ell_2$-norm. Applying stop-gradient (\texttt{stopgrad}) on the target and averaging over all valid prediction positions yields the NAPE objective:
\begin{equation}
  \mathcal{L}
  \;=\; \frac{1}{N-1} \sum_{t=1}^{N-1}
    \mathcal{D}\!\left(\texttt{stopgrad}(z_{t+1}), \; \hat{z}_{t+1}\right).
  \label{eq:nape-loss}
\end{equation}
The \texttt{stopgrad} operator treats the target embedding as a constant, so gradients flow only through the predicted side. Cosine similarity is magnitude-invariant, which prevents the trivial solution of shrinking both sides of the objective toward zero norm. We compare cosine similarity against alternative similarity functions in Section~\ref{sec:ablations} and find it to yield the optimal results.

\paragraph{Causal Transformer Encoder ($h$).} We use a Vision Transformer backbone~\citep{dosovitskiy2020image, wang2026vit} with pre-norm design and a causal attention mask during pre-training (see Figure~\ref{fig:architecture}, right panel). For stability at depth, we adopt Rotary Position Embedding (RoPE)~\citep{su2024roformer} applied independently along the frequency and time axes, LayerScale~\citep{touvron2021going} on residual branches, and parameter-free query-key normalization~\citep{henry2020query} on the per-head query and key projections. We instantiate three configurations of increasing size to test NAPE's scalability: \emph{Small} ($d=384$, $12$ layers, $6$ heads; $\sim$19M parameters), \emph{Base} ($d=768$, $12$ layers, $12$ heads; $\sim$85M parameters), and \emph{Large} ($d=1024$, $24$ layers, $16$ heads; $\sim$303M parameters).

\paragraph{Predictor Head ($g$).} The predictor head $g$ decouples the representation space of $h$ from the space in which predictions are made. Predicting directly with the encoder output forces the encoder to place its representations in the same space as the prediction targets, which can restrict feature richness. This asymmetric predictor design, together with the stop-gradient applied to the target branch, gives NAPE a structural resemblance to SimSiam~\citep{chen2021simsiam}: representation collapse is avoided without contrastive negatives or an EMA teacher, relying instead on the asymmetry between the two branches. NAPE differs from SimSiam in what the branches encode: SimSiam compares two augmented views of the same input under a symmetric encoder, whereas NAPE compares a prediction of the next patch to that patch's actual embedding within a single autoregressive sequence, replacing view augmentation and siamese symmetry with temporal prediction as the source of the learning signal. We study multiple predictor styles in Section~\ref{sec:ablations}.

\begin{wrapfigure}{r}{0.47\textwidth}
\vspace{-2.2em}
\begin{minipage}{0.45\textwidth}
\begin{algorithm}[H]
\caption{NAPE pre-training algorithm.}
\label{alg:nape}
\begin{lstlisting}[style=pseudocode,linewidth=0.46\textwidth]
#f: Patch Embedding Layer
#h: Causal Transformer Encoder
#g: Predictor

for x in loader:       # x: [B,1,F,T_frames]
    z = f(x)           # embeddings [B,T,D]
    z_hat = g(h(z))    # predictions [B,T,D]
    loss = D(z, z_hat)
    loss.backward();   # update(f,h,g)
def D(z, z_hat):
    target = z[:,1:, :].detach() # stop-grad
    pred = z_hat[:,:-1, :]       # AR shift
    pred = normalize(pred, dim=-1)
    target = normalize(target, dim=-1)
    return -(pred*target).sum(-1).mean()
\end{lstlisting}
\end{algorithm}
\end{minipage}
\vspace{-1em}
\end{wrapfigure}

\paragraph{NAPE's Key Components.} Three complementary mechanisms prevent NAPE from converging to trivial solutions and together define its training regime. (i) \emph{\textbf{causality}}: enforced by the causal attention mask that restricts each position to attend only to prior patches, it prevents the encoder from attending to the target patch when producing its prediction, blocking the trivial identity mapping. (ii) The \emph{\textbf{prediction shift}} between input and target positions ensures that the model at position $t$ predicts the embedding at position $t+1$ rather than the embedding at its own position, so causality alone cannot be side-stepped by copying the current input through. (iii) \emph{\textbf{stop-gradient}} on the target embedding prevents gradients from flowing through both sides of the loss simultaneously, which would otherwise allow the encoder to collapse all embeddings toward a shared constant. Together, these three ingredients constitute the core of NAPE's self-supervised recipe, and we analyze the individual contribution of each in Section~\ref{sec:ablations}. The pseudocode of NAPE's pre-training is given in Algorithm~\ref{alg:nape}.

%Without this shift, the objective collapses: the model can trivially satisfy the loss by outputting its own input at every position.

%\subsection{Downstream Adaptation}
%\label{sec:nape-downstream}

%We evaluate NAPE on downstream tasks by discarding the predictor head $g$ used during pre-training and attaching a linear classification head on top of the encoder. We consider two adaptation regimes: (i) \emph{full fine-tuning}, where the encoder and classifier are trained jointly, and (ii) \emph{linear probing}, where the encoder is frozen and only the classifier is trained. %Full recipes for both settings are given in Section~\ref{sec:setup}.

\section{Experiments}
\label{sec:experiments}

In this section, we carry out extensive experiments to evaluate NAPE at different scales and configurations. \textbf{(1)} We first perform several ablations on key design choices to converge on the best NAPE configuration (Section~\ref{sec:ablations}); \textbf{(2)} we then compare that configuration against state-of-the-art methods at multiple scales (Section~\ref{sec:scaling} and ~\ref{sec:sota}); \textbf{(3)} we assess linear separability of the learned features (Section~\ref{sec:linear-probe}), and \textbf{(4)} we finally analyze how NAPE organizes acoustic information through its attention patterns and the structure of its learned embeddings (Section~\ref{sec:qualitative}). 

\subsection{Experimental Setup}
\label{sec:setup}

\paragraph{Pre-training Data.} We pre-train NAPE on
AudioSet~\citep{gemmeke2017audio} without labels, combining the unbalanced and balanced training splits. We obtained and processed $1{,}964{,}222$ clips from the unbalanced split, $20{,}961$ clips from the balanced split, and $18{,}900$ clips from the evaluation split, consistent with prior work~\citep{chen2022beats, chen2024eat, alex2025sslam}. All input waveforms are resampled to mono at $16$\,kHz and converted into log-mel spectrograms with $128$ mel bands using a $25$\,ms Hanning window and $10$\,ms hop size. A $10$-second clip yields a spectrogram of size $1 \times 128 \times 1008$ (channel, frequency, time), corresponding to $8 \times 63 = 504$ non-overlapping $16 \times 16$ patches per clip (as in ~\citep{gong2021ast, chen2022beats, chen2024eat}).

\paragraph{Downstream Benchmarks.} For downstream evaluation, we fine-tune on \textbf{AudioSet-2M} (\textit{AS-2M}, unbalanced) and \textbf{AudioSet-20K} (\emph{AS-20K}, balanced), applying the weighted sampling strategy of~\citet{huang2022masked} on AS-2M to mitigate class imbalance. We further evaluate on
\textbf{ESC-50}~\citep{piczak2015esc}, a $50$-class environmental sound classification benchmark of $2000$ clips at $5$s each, using $5$-fold cross-validation; \textbf{Speech Commands V1 and V2} (\emph{KS1}, \emph{KS2})~\citep{warden2018speech}, keyword spotting tasks with $12$ and $35$ classes respectively; and \textbf{IEMOCAP} (\emph{ER})~\citep{busso2008iemocap}, a $4$-class speech emotion recognition benchmark with $5$-fold cross-validation. Together these tasks span both audio and speech domains. We report mean average precision (mAP) on the multi-label AudioSet tasks and top-1 classification accuracy on the single-label tasks; for ESC-50 and IEMOCAP we report the mean across cross-validation folds.

\paragraph{Pre-training Details.} We pre-train with the AdamW optimizer~\citep{loshchilov2017decoupled} using a base learning rate of $5 \times 10^{-3}$ with cosine decay, a weight decay of $0.05$, $\beta_1 = 0.9$, $\beta_2 = 0.95$, and a batch size of $256$ for the Small and Base configurations, and $128$ for NAPE Large. We pre-train NAPE Small and NAPE Large for $25$ epochs and NAPE Base for $30$ epochs, with a warmup ratio of $10\%$. All models are trained using HuggingFace \texttt{Trainer} with distributed data parallelism (DDP) on NVIDIA L40s GPUs (46GB).

\paragraph{Fine-tuning Details.} For each downstream task, we initialize from the pre-trained encoder and attach a linear classifier on top of mean-pooled patch tokens. We disable the causal attention mask during fine-tuning so that attention is bidirectional over the full patch sequence. We provide an ablation study on the optimal pooling method and whether to use causal/bidirectional attention in the Appendix~\ref{sec:additional_ablations}. We use AdamW with $\beta_2 = 0.999$, cosine learning-rate decay, and layer-wise learning-rate decay~\citep{clark2020electra, bao2021beit}. We use binary cross-entropy for AudioSet and soft-target cross-entropy for the single-label tasks. Following prior work~\citep{gong2021ast,chen2024eat,chen2022beats}, we apply a standard augmentation stack (SpecAugment~\citep{park2019specaugment}, Mixup~\citep{zhang2017mixup}, CutMix~\citep{yun2019cutmix},
DropPath~\citep{huang2016deep}, temporal roll, additive noise, and label smoothing~\citep{szegedy2016rethinking}). For AS-2M and AS-20K we additionally maintain an exponential moving average (EMA)~\citep{polyak1992acceleration} of the fine-tuning weights (decay $0.99995$ and $0.999$ respectively); no EMA is used for the other tasks. Full task-specific hyperparameters are reported in the Appendix~\ref{sec:hyperparameters}. 

\paragraph{Linear Probing Details.} For linear probing, we freeze the pre-trained encoder and train only the classifier plus a preceding LayerNorm~\citep{ba2016layer}. All augmentations are disabled. %Following~\citet{skean2025layer, bolya2026perception}, we conduct linear probing experiments using representations from different layers to find out the layer yielding the best downstream performance.

\paragraph{Baselines.} We primarily compare against recent in-domain self-supervised methods, including Audio-MAE~\citep{huang2022masked}, BEATs~\citep{chen2022beats}, A-JEPA~\citep{fei2023jepa}, ASiT~\citep{ahmed2024asit}, EAT~\citep{chen2024eat}, SSLAM~\citep{alex2025sslam}, and SPEAR~\citep{yang2025spear}. We also include results from out-of-domain and in-domain supervised pre-training. For all baseline methods, fine-tuning results are taken from the original papers.

\subsection{NAPE's Optimal Configuration}
\label{sec:ablations}
In this section, we ablate the main design axes of NAPE to identify its optimal configuration. Unless otherwise specified, all ablations use the NAPE \textit{Base} model and the raster scanning order. Each ablation modifies a single design axis at a time, holding the rest of the configuration fixed at NAPE's defaults.

\paragraph{\hlc{green_ablation}{Prediction Shift/Stop-gradient/Causality.}} Table~\ref{tab:maincomponents} disentangles the three key mechanisms of NEPA: the \emph{prediction shift}, the \emph{stop-gradient} on the target, and the \emph{causal attention mask}. Removing either the prediction shift or the stop-gradient causes
pre-training to diverge, matching the analysis in
Section~\ref{sec:nape-prediction}: without the shift, the model at position $t$ can trivially satisfy the objective by copying its own input embedding through; without the stop-gradient, gradients flow through both sides of the loss and the encoder collapses all embeddings toward a shared constant.

\begin{wraptable}{r}{0.54\textwidth}
\small
%\renewcommand{\arraystretch}{0.9}
%\vspace{-1em}
\renewcommand{\tabcolsep}{0.5mm}
\centering
\caption{\hlc{green_ablation}{\textbf{Ablation on main NAPE's design elements}}: \emph{prediction shift}, \emph{stop-gradient}, and \emph{causal objective}.}
\label{tab:maincomponents}
\begin{tabular}{ccccccccc}
\toprule
\textbf{Pred} & \textbf{stop} & \textbf{causal} & \multicolumn{3}{c}{\textbf{Audio Tasks}}  & \multicolumn{3}{c}{\textbf{Speech Tasks}}\\
\textbf{shift} & \textbf{grad} & \textbf{mask} & \emph{AS-2M} & \emph{AS-20K} & \emph{ESC-50} & \emph{KS1} & \emph{KS2} & \emph{ER} \\
\midrule
\ding{56} & \ding{51} & \ding{51} & \multicolumn{6}{c}{\color{red}\textbf{Diverge}} \\
\ding{51} & \ding{56} & \ding{51} & \multicolumn{6}{c}{\color{red}\textbf{Diverge}} \\
\hline
\ding{51} & \ding{51} & \ding{56} & 41.8 & 24.8 & 68.9 & 96.1 & 97.3 & 57.0 \\
\ding{51} & \ding{51} & \ding{51} & \textbf{49.6} & \textbf{39.1} & \textbf{94.2} &\textbf{97.9} & \textbf{98.8} & \textbf{64.9} \\
\bottomrule
\end{tabular}
\vspace{-1em}
\end{wraptable}

Removing the causal mask, in contrast, does not diverge: the
prediction shift alone continues to define a nominal prediction target, but degrades downstream performance sharply, with the largest drops on the audio benchmarks ($-7.8$ mAP on AS-2M, $-14.3$ on AS-20K, and $-25.4$ points on ESC-50; smaller but consistent drops on the speech tasks). Without causality, the encoder can attend to the target patch while producing its prediction, so the objective is trivially satisfied by a near-identity mapping that routes each target back to its predicted position (the loss saturates near $-1$ within a few thousand steps of pre-training, see Appendix~\ref{sec:loss_visualization}): the loss decreases during pre-training, but the encoder is not forced to learn useful structure. All three mechanisms are therefore jointly necessary, none can be dropped without either destabilizing training or degrading the learned representations to a degree that fine-tuning cannot recover.

\begin{wraptable}{r}{0.49\textwidth}
\small
\vspace{-1.5em}
\renewcommand{\tabcolsep}{0.5mm}
\centering
    \caption{\hlc{orange_ablation}{Ablation on the patch embedding layer.}}
%\resizebox{0.999\linewidth}{!}{
\begin{tabular}{lcccccc}
\toprule
\multicolumn{1}{c}{\textbf{Patch Emb.}} & \multicolumn{3}{c}{\textbf{Audio Tasks}}  & \multicolumn{3}{c}{\textbf{Speech Tasks}}\\ \multicolumn{1}{c}{\textbf{Layer}} & \emph{AS-2M} & \emph{AS-20K} & \emph{ESC-50} & \emph{KS1} & \emph{KS2} &\emph{ER} \\ 

\midrule
Convstem & 46.7 & 34.3 & 89.1 & 97.4 & 98.3 & 63.6 \\
Speechstem & 47.6 & 33.1 & 88.4 & \textbf{98.1} & \textbf{98.9} & 63.0 \\
\hline \addlinespace[2pt]
\textbf{Conv2d} & \textbf{49.6} & \textbf{39.1} & \textbf{94.2} & 97.9 & 98.8 & \textbf{64.9} \\

\bottomrule
\end{tabular}%}
\label{tab:patchemb}
\vspace{-1em}
\end{wraptable}

\paragraph{\hlc{orange_ablation}{Patch Embedding Layer $f$.}} Table~\ref{tab:patchemb} compares three patchifiers, all configured to produce the same sequence length (504 tokens per $10$s clip). The default \emph{Conv2d} is a single strided convolution with kernel and stride $16 \times 16$, which treats time and frequency as symmetric 2D axes and matches the standard ViT design~\citep{dosovitskiy2020image}. We also consider two alternatives motivated by observations from computer vision and speech literature. The \emph{Convstem}, following \citet{xiao2021early}, replaces the single Conv2d with four $3 \times 3$ stride-2 convolutions interleaved with batch normalization with the channel count growing progressively toward $d$; deeper convolutional stems have been shown to stabilize optimization and improve downstream performance in ViT-based image models. The \emph{Speechstem}, inspired by the audio front-end of speech-oriented models such as Gemma~3n and Gemma 4~\citep{team2026gemma}, flattens the mel axis into feature channels and applies temporal-only $3 \times 3$ convolutions with a temporal downsampling factor of $2$, producing a 1D time-only sequence of $504$ tokens that reflects the frame-based processing typical of speech recognition systems.

Both alternatives underperform Conv2d across all benchmarks, with the largest gaps on AudioSet: on AS-20K, Convstem loses $4.8$ mAP points and Speechstem loses $6.0$ mAP spoints relative to Conv2d. Convstem's added non-linearity and speechstem's temporal-first inductive bias therefore do not translate into gains for our causal spectrogram prediction objective, treating frequency and time as symmetric 2D axes at the patchification stage is important for the pretraining signal that NAPE exploits. Conv2d remains NAPE's default.

\begin{wraptable}{r}{0.555\textwidth}
\small
\vspace{-1.5em}
\renewcommand{\tabcolsep}{0.5mm}
\centering
    \caption{\hlc{yellow_ablation}{Ablation on the predictor-style variants.}}
%\resizebox{0.999\linewidth}{!}{

\begin{tabular}{lccccccc}
\toprule
\multicolumn{1}{c}{\textbf{Predictor}} & \multirow{2}{*}{\textbf{\#Par.}} & \multicolumn{3}{c}{\textbf{Audio Tasks}}  & \multicolumn{3}{c}{\textbf{Speech Tasks}}\\ \multicolumn{1}{c}{\textbf{Style}} & & \emph{AS-2M} & \emph{AS-20K} & \emph{ESC-50} & \emph{KS1} & \emph{KS2} &\emph{ER} \\ 

\midrule

\textcolor{darkgray}{None} & - & 48.7 & 37.8 & 93.3 & 98.1 & \textbf{98.8} & 64.2 \\
2-MLP & 1.2M & 49.4 & 38.5 & 93.6 & 98.0 & \textbf{98.8} & 64.2 \\
Transformer & 14.2M & 49.2 & 38.4 & 93.0 & \textbf{98.2} & 98.7 & \textbf{65.0} \\
\textbf{SimSiam} & 1.8M & \textbf{49.6} & \textbf{39.1} & \textbf{94.2} & 97.9 & \textbf{98.8} & 64.9 \\

\bottomrule
\end{tabular}%}
\label{tab:predictor_table}
\vspace{-1em}
\end{wraptable}

\paragraph{\hlc{yellow_ablation}{Predictor $g$.}} 
Table~\ref{tab:predictor_table} compares four predictor variants: no predictor (encoder output used directly as
the prediction), a two-layer MLP (\emph{2-MLP};
$\text{Linear}(d,d) \to \text{GELU} \to \text{Linear}(d,d)$),
a \textit{SimSiam}-style~\citep{chen2021simsiam} three-layer predictor with intermediate LayerNorms ($\text{Linear}(d,d) \to \text{LN} \to \text{GELU} \to \text{Linear}(d,d) \to \text{LN} \to \text{GELU} \to \text{Linear}(d,d)$),
and a \emph{Transformer} predictor in the style of JEPA-family methods: a 2-layer causal Transformer ($16$ heads) operating in the encoder's hidden dimension (we apply a causal mask to prevent future leaking as for the encoder). %The causal mask is essential for the Transformer predictor — without it, position $t$ could attend to position $t+1$ and directly read the target it is supposed to predict, recreating the same target-leakage failure mode observed when the encoder's causal mask is removed. 
Adding any predictor improves over using the encoder output directly, confirming that decoupling the representation and prediction spaces benefits the learned features. Among the three predictor styles, the SimSiam variant performs best on the audio benchmarks and on IEMOCAP, while remaining on par on the keyword-spotting tasks. Notably, the lightweight SimSiam predictor outperforms the Transformer predictor despite being nearly $8\times$ smaller in parameters, indicating that additional predictor capacity is not the bottleneck for NAPE, consistent with the observation in the siamese self-supervised setting~\citep{chen2021simsiam} that a compact MLP predictor suffices when combined with a well-designed encoder and stop-gradient target. The SimSiam-style predictor is NAPE's default.

\begin{figure}[t]
    \centering
    \includegraphics[width=1\textwidth]{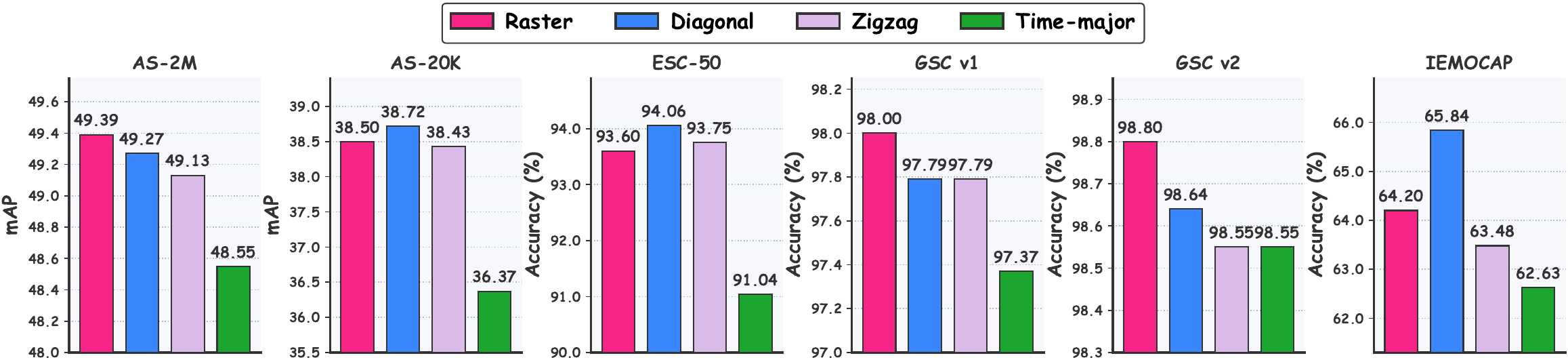}
    \caption{\hlc{blue_ablation}{NAPE's performance across four scan orders on six benchmarks.}}
    \label{fig:variants}
    %\vspace{-0.3cm}
\end{figure}

\paragraph{\hlc{blue_ablation}{Patch Scannning Variants.}} Figure~\ref{fig:variants} compares the four scanning orders introduced in Section~\ref{sec:nape-patches} (we use the 2-MLP predictor style). \emph{Diagonal}, \emph{raster}, and \emph{zigzag} all perform comparably well across the six benchmarks, with diagonal and raster slightly ahead of zigzag on most tasks. In contrast, \emph{time-major} consistently underperforms the other three orders across all tasks. We interpret this pattern as reflecting the temporal structure of audio: diagonal, raster, and zigzag all advance in time as the causal sequence progresses, allowing the model to accumulate temporal context in a way that matches how acoustic events unfold. Time-major, instead, exhausts each frequency column before advancing in time, so predictions early in the sequence are conditioned on rich instantaneous spectra but only limited temporal context, a mismatch with the temporal nature of audio classification. Given their superior performance, we retain both raster and diagonal in the main comparison against state-of-the-art methods in Section~\ref{sec:sota}.

\begin{wraptable}{r}{0.505\textwidth}
\small
\vspace{-1.5em}
\renewcommand{\tabcolsep}{0.5mm}
\centering
    \caption{\hlc{pink_ablation}{Ablation on NAPE's target to predict.}}
%\resizebox{0.999\linewidth}{!}{
\begin{tabular}{lcccccc}
\toprule
\multicolumn{1}{c}{\textbf{Predicted}} & \multicolumn{3}{c}{\textbf{Audio Tasks}}  & \multicolumn{3}{c}{\textbf{Speech Tasks}}\\ \multicolumn{1}{c}{\textbf{Target}} & \emph{AS-2M} & \emph{AS-20K} & \emph{ESC-50} & \emph{KS1} & \emph{KS2} &\emph{ER} \\ 

\midrule
1st enc. layer & \multicolumn{6}{c}{\color{red}\textbf{Diverge}}  \\
\hline \addlinespace[2pt]
Raw Mel & \textbf{49.7} & 38.0 & \textbf{94.8} & 97.7 & 98.6 & 64.2 \\
\textbf{Patch embed} & 49.6 & \textbf{39.1} & 94.2 & \textbf{97.9} & \textbf{98.8} & \textbf{64.9} \\

\bottomrule
\end{tabular}%}
\label{tab:target}
\vspace{-1em}
\end{wraptable}
\paragraph{\hlc{pink_ablation}{Prediction Target.}} Table~\ref{tab:target} compares three choices for the target of the auto-regressive prediction: (i) the \emph{patch embedding} $z_{t+1}$ produced by the shared embedding layer $f$ (the default); (ii) the \emph{raw mel} content of the next patch i.e., the flattened mel-spectrogram values inside the patch, following the target formulation used by masked reconstruction methods such as Audio MAE~\citep{huang2022masked}; (iii) the output of the \emph{first encoder layer}, in the style of JEPA methods that use deeper encoder features as prediction targets. For the raw mel variant, we add a linear projection on top of the predictor $g$ to map its output from $d$ dimensions to the raw patch dimensionality.

Patch embedding and raw mel yield comparable results across all benchmarks, indicating that both are valid target choices for NAPE. Using the first encoder layer as target, however, causes pre-training to diverge: the target itself depends on the encoder being trained, and stop-gradient alone is insufficient to prevent the encoder from collapsing both sides of the loss to a shared constant. JEPA-family methods circumvent this instability with additional regularization such as EMA teachers~\citep{assran2023self, fei2023jepa}, variance-covariance regularizers such as VISReg~\citep{wu2026visreg} and sketched isotropic gaussian regularizers such as SIGReg~\citep{balestriero2025lejepa}, which we do not employ here. Since the patch embedding approach on average performs better than raw mel and it is adopted in ~\citep{xu2025nepa} as well, we retain the patch embedding as NAPE's default.

\begin{wraptable}{r}{0.5\textwidth}
\small
\vspace{-1.5em}
\renewcommand{\tabcolsep}{0.5mm}
\centering
    \caption{\hlc{violetblue_ablation}{Ablation on the similarity function.}}
%\resizebox{0.999\linewidth}{!}{
\begin{tabular}{lcccccc}
\toprule
\multicolumn{1}{c}{\textbf{Similarity}} & \multicolumn{3}{c}{\textbf{Audio Tasks}}  & \multicolumn{3}{c}{\textbf{Speech Tasks}}\\ \multicolumn{1}{c}{\textbf{Function}} & \emph{AS-2M} & \emph{AS-20K} & \emph{ESC-50} & \emph{KS1} & \emph{KS2} &\emph{ER} \\ 

\midrule

L1 & \multicolumn{6}{c}{\color{red}\textbf{Diverge}}  \\
L2 & \multicolumn{6}{c}{\color{red}\textbf{Diverge}} \\
\hline \addlinespace[2pt]
Cross-entropy & 48.8 & 37.4 & 93.6 & \textbf{98.0} & 98.7 & 64.5 \\
\textbf{Cosine} & \textbf{49.6} & \textbf{39.1} & \textbf{94.2} & 97.9 & \textbf{98.8} & \textbf{64.9} \\

\bottomrule
\end{tabular}%}
\label{tab:loss}
\vspace{-1em}
\end{wraptable}

\paragraph{\hlc{violetblue_ablation}{Similarity Function ($\mathcal{D}$).}} In Table~\ref{tab:loss} we compare four choices for the similarity function $\mathcal{D}$ in Eq.~\ref{eq:cosine-similarity}: the \emph{negative cosine similarity} (NAPE's default), a soft \emph{cross-entropy} formulation that treats prediction and target as distributions over the $d$ channels after applying a softmax (as in~\cite{chen2021simsiam}), and the $\ell_1$ and $\ell_2$ distances between the two $d$-dimensional vectors. Both $\ell_1$ and $\ell_2$ cause pre-training to diverge: unlike cosine similarity, which is magnitude-invariant, these distance-based objectives can be trivially minimized by shrinking the norm of both predicted and target embeddings toward zero, a form of representation collapse in which the encoder outputs converge to a shared low-magnitude constant. The cross-entropy variant trains stably (softmax normalization implicitly bounds the target magnitude) but underperforms cosine on all benchmarks except the keyword-spotting tasks, where the two are on par. Cosine similarity, combining magnitude-invariance with a directionally informative loss signal, yields the best overall results and is retained as NAPE's default.

\begin{boxaccent}{NAPE's Optimal Configuration}
(i) \textbf{Conv2d} patch embedding layer, (ii) \textbf{raster}/\textbf{diagonal} scanning order, (iii) \textbf{SimSiam-style predictor} with three-layer MLP and intermediate LayerNorms, (iv) \textbf{patch embedding} as the prediction target, and (v) \textbf{negative cosine similarity} loss with stop-gradient on the target branch.
\end{boxaccent}

\textbf{Additional Ablation Results.} We refer to Appendix~\ref{sec:additional_ablations} for additional ablations studies on: \textbf{1)} the use of normalization layers (LayerNorm vs RMSNorm), \textbf{2)} freezing/unfreezing the patch embedding layer, \textbf{3)} the optimal positional encoding (absolute encoding vs RoPE), \textbf{4)} the optimal pooling method during fine-tuning, \textbf{5)} the optimal attention type during fine-tuning (causal vs bidirectional), \textbf{(6)} the additional use of random masking during pre-training, and \textbf{7)} NAPE's performance in terms of different compute budgets.

\subsection{Scaling NAPE}
\label{sec:scaling}

\begin{wrapfigure}{r}{0.45\textwidth}
    \centering
    \vspace{-3em}
    \includegraphics[width=\linewidth]{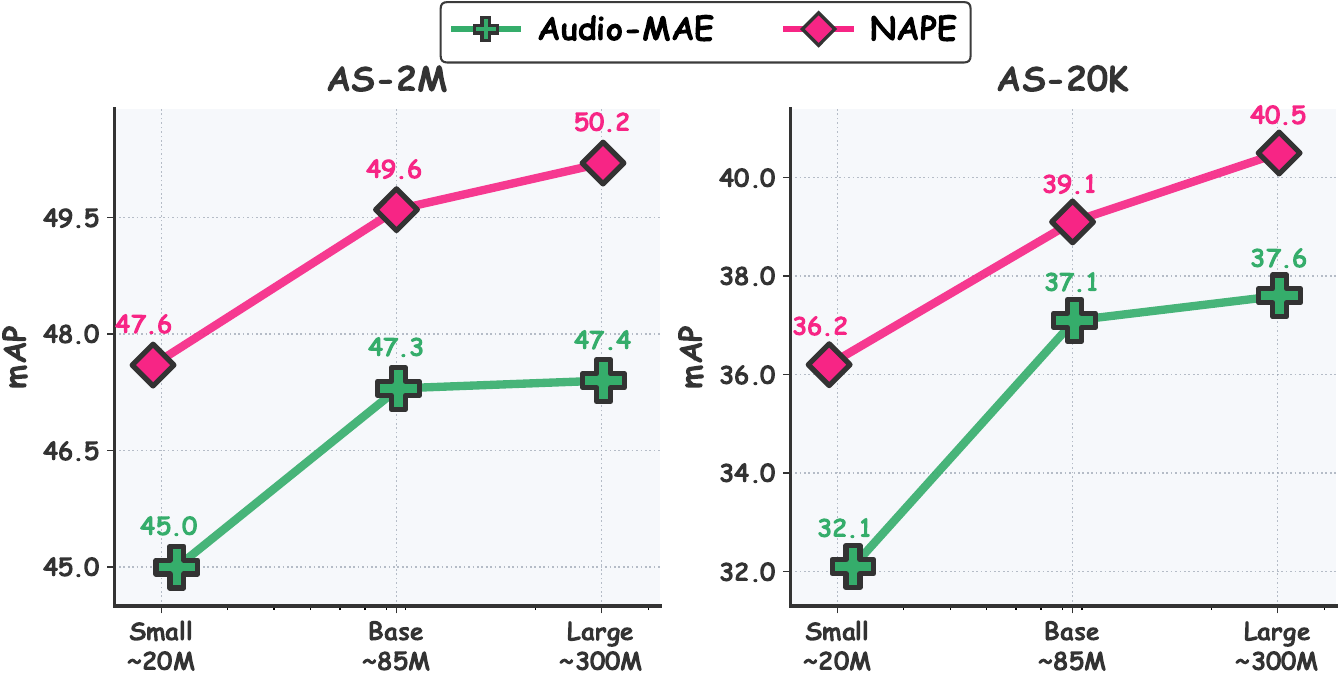}
    \caption{Scaling comparison between Audio-MAE and NAPE, raster.}
    \label{fig:audiomae_vs_nape}
    \vspace{-1em}
\end{wrapfigure}

\begin{figure}[!t]
    \centering
    \includegraphics[width=1\textwidth]{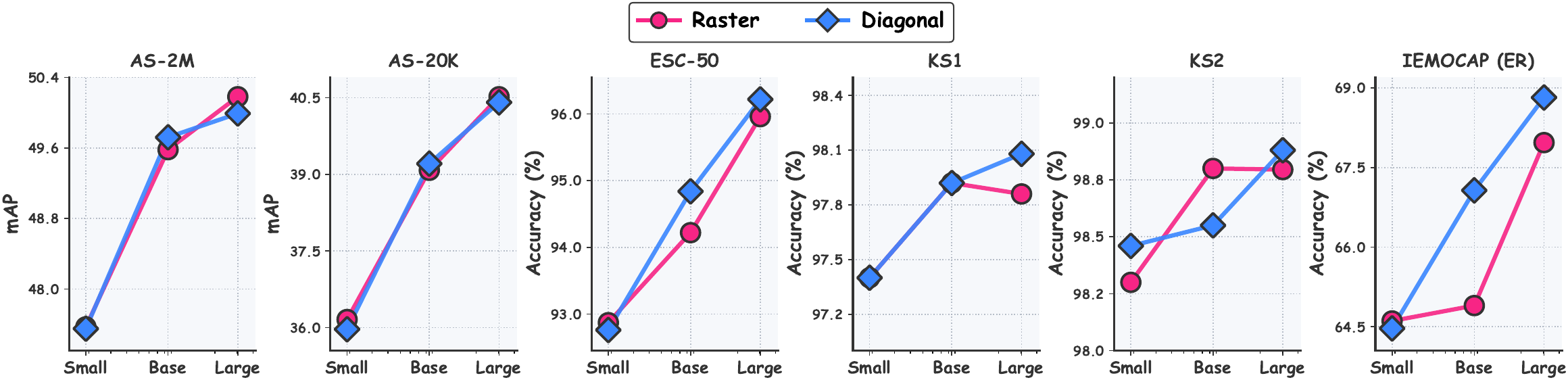}
    \caption{NAPE's results at different scales under raster and diagonal scan orders.}
    \label{fig:main_results}
    \vspace{-0.5cm}
\end{figure}

Figure~\ref{fig:main_results} shows downstream performance for the raster and diagonal variants of NAPE at three encoder scales: \emph{Small} ($\sim$19M parameters, NAPE-S), \emph{Base} ($\sim$85M, NAPE-B), and \emph{Large} ($\sim$303M, NAPE-L). NAPE scales positively across the board: on every one of the six benchmarks, NAPE-L improves over NAPE-B, which in turn improves over the small version. While this improvement tends to diminish as we scale from the base to the large model, we observe consistent scaling gains for most of the tasks, while the keyword-spotting tasks show smaller absolute gains, reflecting their already-saturated accuracy levels. Raster and diagonal track each other closely at the Small scale, with essentially identical performance across all six tasks. At the Base and Large scales, diagonal slightly outperforms raster on most benchmarks, while raster achieves the strongest result on AS-2M with NAPE-L reaching $50.18$ mAP. %Overall, NAPE exhibits favorable scaling trends across both audio and speech benchmarks, supporting autoregressive next patch embedding prediction as a scalable objective for audio representation learning.
In Figure~\ref{fig:audiomae_vs_nape}, we compares NAPE (raster) and Audio-MAE~\citep{huang2022masked} at three encoder scales on AS-2M and AS-20K. NAPE outperforms Audio-MAE at every scale, with particularly large margins at the Small size ($+2.6$ mAP on AS-2M, $+4.1$ mAP on AS-20K). NAPE also
exhibits more favorable scaling behavior from Base to Large.

\begin{table}[t]
\small
\renewcommand{\tabcolsep}{1.5mm}
\centering
    \caption{Comparison with audio methods on audio and speech downstream tasks. IN, AS,
and LS denote the ImageNet, AudioSet, and LibriSpeech datasets, respectively. TI denotes
the 400M text-image pairs for CLIP pre-training. We \textcolor{darkgray}{gray-out} the models and results with additional supervised training on external datasets. Best results are in \textbf{bold}, second-best are \underline{underlined}.}
%\resizebox{0.999\linewidth}{!}{
\begin{tabular}{lcccccccc}
\toprule
\multirow{2}{*}{\textbf{Model}} & \multirow{2}{*}{\textbf{\#Par.}} & \textbf{Pre-train} & \multicolumn{3}{c}{\textbf{Audio Tasks}}  & \multicolumn{3}{c}{\textbf{Speech Tasks}}\\ & & \textbf{Data} & \emph{AS-2M} & \emph{AS-20K} & \emph{ESC-50} & \emph{KS1} & \emph{KS2} &\emph{ER} \\

\midrule
 \multicolumn{3}{l}{\textit{\textbf{Out-of-domain Supervised Pre-training}}} & & & & & \\ \addlinespace[2pt] 
\textcolor{darkgray}{PSLA} \citep{gong2021psla} & \textcolor{darkgray}{14M} & \textcolor{darkgray}{IN} & \textcolor{darkgray}{44.4} & \textcolor{darkgray}{31.9} & \textcolor{darkgray}{-}& \textcolor{darkgray}{-}& \textcolor{darkgray}{96.3} & \textcolor{darkgray}{-}\\ 
\textcolor{darkgray}{AST} \citep{gong2021ast} & \textcolor{darkgray}{86M} & \textcolor{darkgray}{IN} & \textcolor{darkgray}{45.9} & \textcolor{darkgray}{34.7} & \textcolor{darkgray}{88.7} & \textcolor{darkgray}{95.5} & \textcolor{darkgray}{98.1} & \textcolor{darkgray}{56.0} \\ 
\textcolor{darkgray}{HTS-AT} \citep{chen2022hts} & \textcolor{darkgray}{31M} & \textcolor{darkgray}{IN} & \textcolor{darkgray}{47.1} & \textcolor{darkgray}{-} & \textcolor{darkgray}{-} & \textcolor{darkgray}{-} & \textcolor{darkgray}{98.0} & \textcolor{darkgray}{-} \\
\textcolor{darkgray}{Audio-CLIP} \citep{guzhov2022audioclip} & \textcolor{darkgray}{93M} & \textcolor{darkgray}{TI+AS} & \textcolor{darkgray}{25.9} & \textcolor{darkgray}{-} & \textcolor{darkgray}{96.7} & \textcolor{darkgray}{-} & \textcolor{darkgray}{-} & \textcolor{darkgray}{-} \\

\addlinespace[2pt]

\hline \addlinespace[2pt] 

\multicolumn{3}{l}{\textit{\textbf{In-domain Supervised Pre-training}}} & & & & & \\ \addlinespace[2pt] 

\textcolor{darkgray}{AST} \citep{gong2021ast} & \textcolor{darkgray}{86M} & \textcolor{darkgray}{IN+AS} & \textcolor{darkgray}{45.9} & \textcolor{darkgray}{-} & \textcolor{darkgray}{95.6} & \textcolor{darkgray}{-} & \textcolor{darkgray}{97.9} & \textcolor{darkgray}{-} \\ 
\textcolor{darkgray}{HTS-AT} \citep{chen2022hts} & \textcolor{darkgray}{31M} & \textcolor{darkgray}{IN+AS} & \textcolor{darkgray}{47.1} & \textcolor{darkgray}{-} & \textcolor{darkgray}{97.0} & \textcolor{darkgray}{-} & \textcolor{darkgray}{-} & \textcolor{darkgray}{-} \\
\textcolor{darkgray}{Audio-MAE} \citep{huang2022masked} & \textcolor{darkgray}{86M} & \textcolor{darkgray}{AS} & \textcolor{darkgray}{-} & \textcolor{darkgray}{-} & \textcolor{darkgray}{97.4} & \textcolor{darkgray}{-} & \textcolor{darkgray}{-} & \textcolor{darkgray}{-}\\

\addlinespace[2pt]
\hline \addlinespace[2pt] 

\multicolumn{3}{l}{\textit{\textbf{Self-Supervised Pre-training}}} & & & & & \\ \addlinespace[2pt] 

SS-AST \citep{gong2022ssast} & 89M & AS+LS & - & 31.0 & 88.8 & 96.0 & 98.0 & 59.6 \\
MAE-AST \citep{baade2022mae} & 86M & AS+LS & - & 30.6 & 90.0 & 95.8 & 97.9 & 59.8 \\
CAV-MAE \citep{gong2022contrastive} & 86M & IN+AS & 44.9 & 34.2 & - & - & - & - \\
Audio-MAE \citep{huang2022masked} & 86M & AS & 47.3 & 37.1 & 94.1 & 96.9 & 98.3 & - \\
Audio-MAE L \citep{huang2022masked} & 304M & AS & 47.4 & 37.7 & - & - & - & - \\
data2vec \citep{baevski2022data2vec} & 94M & AS & - & 34.5 & - & - & - & - \\
MaskSpec \citep{chong2023masked} & 86M & AS & 47.1 & 32.3 & 89.6 & - & 97.7 & - \\
BEATs$_{iter3}$ \citep{chen2022beats} & 90M & AS & 48.0 & 38.3 & 95.6 & 97.7 & 98.3 & 64.5 \\
A-JEPA \citep{fei2023jepa} & 86M & AS & 48.6 & 38.4 & \textbf{96.3} & 97.7 & 98.5 & - \\
ASiT \citep{ahmed2024asit} & 86M & AS & 48.0 & 38.6 & 95.3 & \underline{98.2} & \textbf{98.9} & - \\
EAT \citep{chen2024eat} & 88M & AS & 48.6 & 40.2 & 95.9 & - & 98.3 & - \\
SSLAM \citep{alex2025sslam} & 88M & AS & \textbf{50.2} & \textbf{40.9} & \underline{96.2} & \textbf{98.8} & 98.1 & - \\
SPEAR$_a$ Large \citep{yang2025spear} & 327M & AS & 49.7 & 39.3 & - & - & - & - \\

\addlinespace[2pt]
\hline \addlinespace[2pt] 

\textbf{\textcolor{nepapink}{NAPE-B raster}} & 85M & AS & 49.6 & 39.1 & 94.2 & 97.9 & \underline{98.8} & 64.9 \\
\textbf{\textcolor{nepadiagonal}{NAPE-B diagonal}} & 85M & AS &49.7 &39.2 &94.8 &97.9 &98.6 &67.1 \\
\textbf{\textcolor{nepapink}{NAPE-L raster}} & 303M & AS &\textbf{50.2} &\underline{40.5} &96.0 &97.9 &\underline{98.8} &\underline{68.0} \\
\textbf{\textcolor{nepadiagonal}{NAPE-L diagonal}} & 303M & AS & \underline{50.0} & 40.4 & \underline{96.2} & \underline{98.2} & \textbf{98.9} & \textbf{68.8}\\
\bottomrule
\end{tabular}%}
\label{tab:main_table}
%\vspace{-0.2cm}
\end{table}

\subsection{Comparison with State-of-the-Art}
\label{sec:sota}

We compare NAPE against prior audio pre-training methods in Table~\ref{tab:main_table}. We use the best configuration identified in Section~\ref{sec:ablations}, reporting both raster and diagonal variants at the Base and Large scales. NAPE-B with diagonal scan delivers strong performance across all six benchmarks, matching or exceeding every self-supervised competitor of comparable size. When we scale it further, NAPE-L raster attains results on par with the strongest baseline, SSLAM~\citep{alex2025sslam}, tying it on AS-2M ($50.2$ mAP) and approaching it closely on AS-20K ($40.5$ vs $40.9$ mAP) and ESC-50 ($96.0$ vs $96.2$\%). What is notable about these results is the simplicity of underlying NAPE's recipe. Unlike SSLAM, which relies on audio-mixture supervision, a student-teacher architecture, and a reconstruction decoder, NAPE requires none of these ingredients: its pre-training objective consists of a stop-gradient and a negative cosine similarity between the predicted next patch embedding and its target. NAPE also outperforms A-JEPA~\citep{fei2023jepa} on $5$ benchmarks despite A-JEPA relying on an auxiliary target encoder updated via EMA, predicting multiple masked patches in parallel rather than causally, and adopting a regularized masking strategy during fine-tuning. Finally, NAPE transfers particularly well to speech tasks: on IEMOCAP, NAPE-L reaches $68.0$\% accuracy, a $+3.5$-point improvement over the strongest baseline result reported at any scale (BEATs$_{iter3}$ at $64.5$\%), suggesting that NAPE  learns representations that generalize well beyond acoustic-event classification.

\subsection{Linear Probing Results}
\label{sec:linear-probe}

\paragraph{\hlc{grey_ablation}{Which Layer to Probe?}} Linear probing measures the linear separability of the features produced by the pretrained encoder, treating the encoder as a fixed feature extractor and training only a linear classification head. Following recent observations that the most classification-relevant features in deep Transformer models often lie in intermediate rather than final layers~\citep{skean2025layer, bolya2026perception}, we begin our linear probing study by measuring downstream performance as a function of encoder depth. \begin{wrapfigure}{r}{0.45\textwidth}
    \centering
    \vspace{-1em}
    \includegraphics[width=\linewidth]{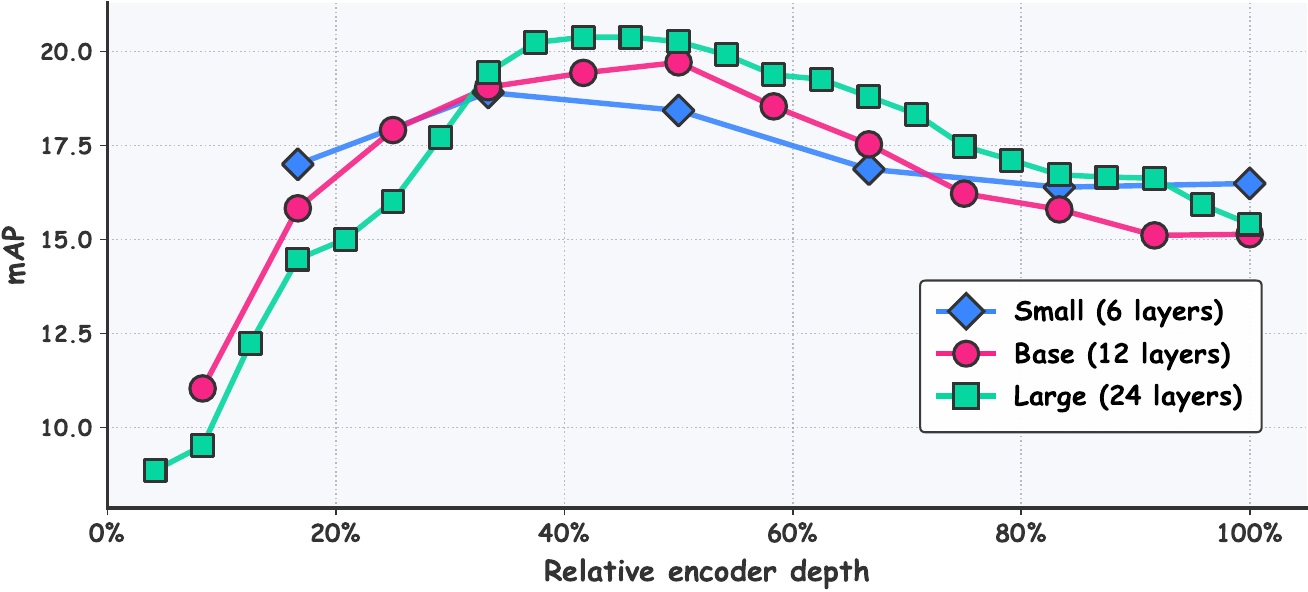}
    \caption{\hlc{grey_ablation}{Layer-wise linear probing analysis.}}
    \label{fig:linearprobing_plot}
    \vspace{-1em}
\end{wrapfigure}Figure~\ref{fig:linearprobing_plot} reports AS-20K mAP obtained by linearly probing each layer of NAPE-S, NAPE-B, and NAPE-L, using the best raster configuration identified in Section~\ref{sec:ablations}. Across all three scales, the best probing layer lies at roughly the middle of the encoder: layer $2$ for NAPE-S, layer $6$ for NAPE-B, and layer $11$ for NAPE-LL. Beyond this mid-network optimum, performance declines steadily toward the final layer, dropping by roughly $3$-$5$ mAP points. This pattern is consistent with the interpretation that the top layers of a NAPE-pretrained encoder specialize for the next-patch-embedding prediction objective, while the mid-layers retain more general and classification-relevant information.

\begin{wraptable}{r}{0.37\textwidth}
\small
\vspace{-0.5cm}
\renewcommand{\tabcolsep}{0.5mm}
\centering
    \caption{\hlc{almondmiddle}{Linear probing results.}}
%\resizebox{0.999\linewidth}{!}{
\begin{tabular}{lcccc}
\toprule
\multicolumn{1}{c}{\textbf{Model}} & \textbf{Layer} & \emph{AS-2M} & \emph{AS-20K} & \emph{ESC-50}\\ 
\midrule
Small & 2nd & 23.2 & 18.9 & 79.8 \\
Base & 6th& 25.0 & 19.7 & 81.7 \\
\textbf{Large} & 11th &\textbf{27.1} &\textbf{20.4} & \textbf{83.5} \\
\bottomrule
\end{tabular}%}
\label{tab:linearprobing_table}
\vspace{-0.5cm}

\end{wraptable}

\paragraph{\hlc{almondmiddle}{Linear Probing Across Scales and Tasks.}} We then use the best probing layer identified per model to compare NAPE-S, NAPE-B, and NAPE-L on AS-2M, AS-20K, and ESC-50. Table~\ref{tab:linearprobing_table} reports the results. NAPE scales positively under linear probing: larger models yield stronger probes on every task, from AS-2M ($+3.9$ mAP from small to large) to ESC-50 ($+3.7$ in accuracy). This trend reinforces the scaling behavior observed under fine-tuning (Section~\ref{sec:scaling}). At the same time, the absolute linear-probing numbers are noticeably lower than their fine-tuning counterparts. This gap is expected for prediction-based self-supervised methods: the pretraining objective encourages the encoder to learn features that support the next-patch prediction task, which do not necessarily align with the linear separability needed for classification.

\subsection{Qualitative Results}
\label{sec:qualitative}

To gain insight into what NAPE learns, we complement the quantitative benchmarks with two qualitative analyses of a pretrained NAPE-L (raster) on the AudioSet evaluation set. Both use the model after pretraining, with no fine-tuning. More qualitative results can be found in the Appendix~\ref{sec:additional_qualitative}.

\begin{figure}[t]
    \centering
    % --- top row: four subfigures ---
    \begin{subfigure}[t]{0.295\textwidth}
        \centering
        \includegraphics[width=\linewidth]{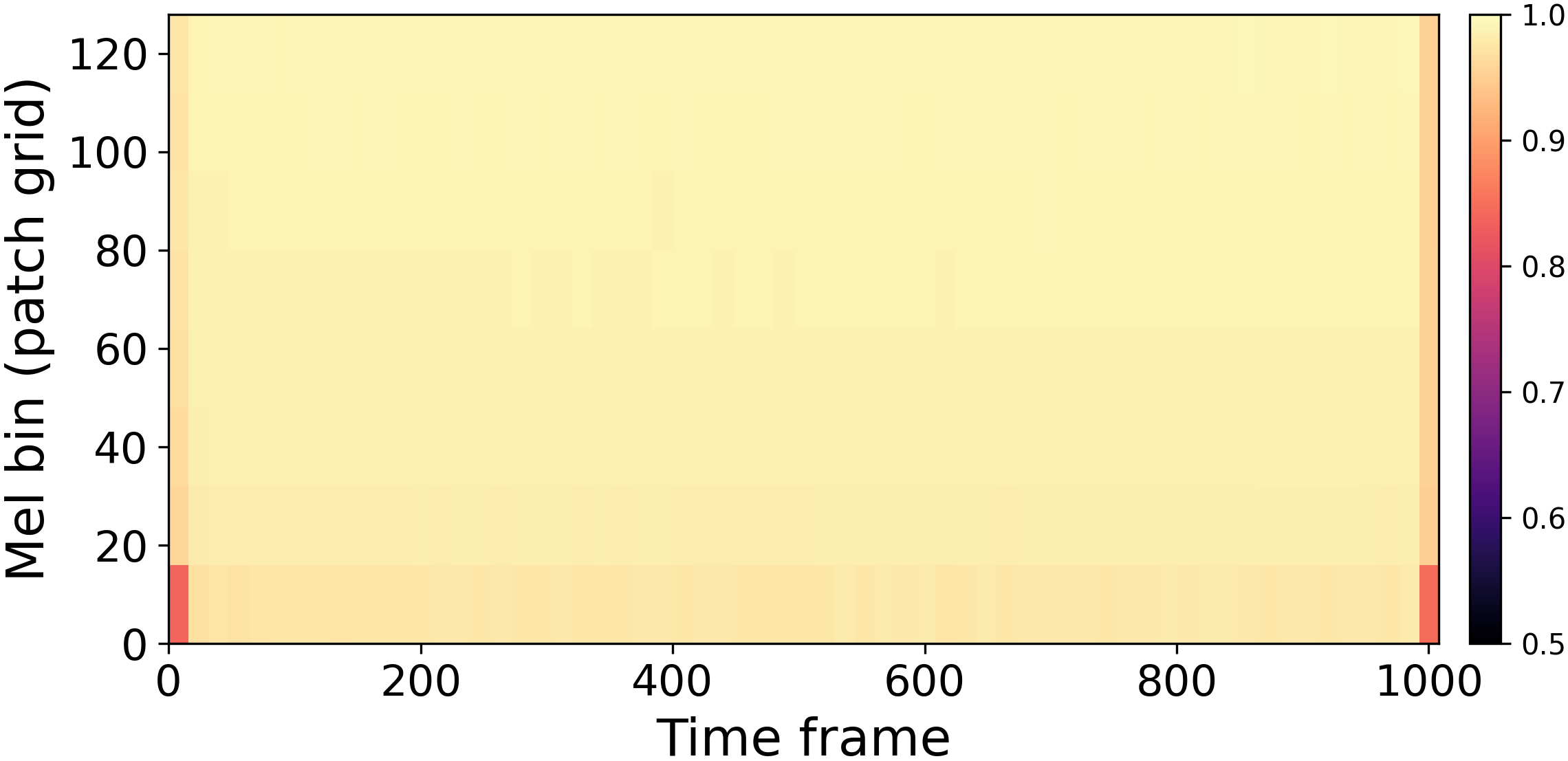}
    \end{subfigure}
    \hfill
    \begin{subfigure}[t]{0.345\textwidth}
        \centering
        \includegraphics[width=\linewidth]{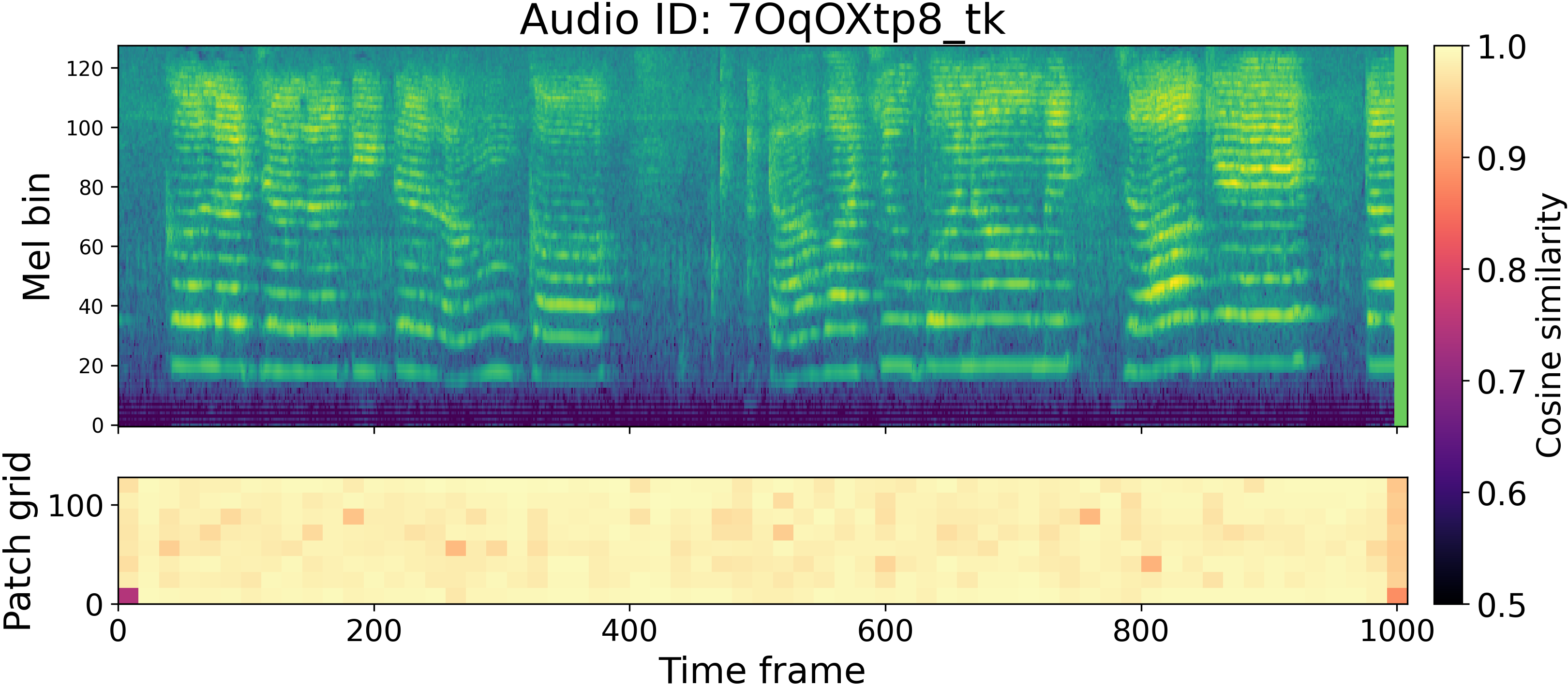}
    \end{subfigure}
    \hfill
    \begin{subfigure}[t]{0.345\textwidth}
        \centering
        \includegraphics[width=\linewidth]{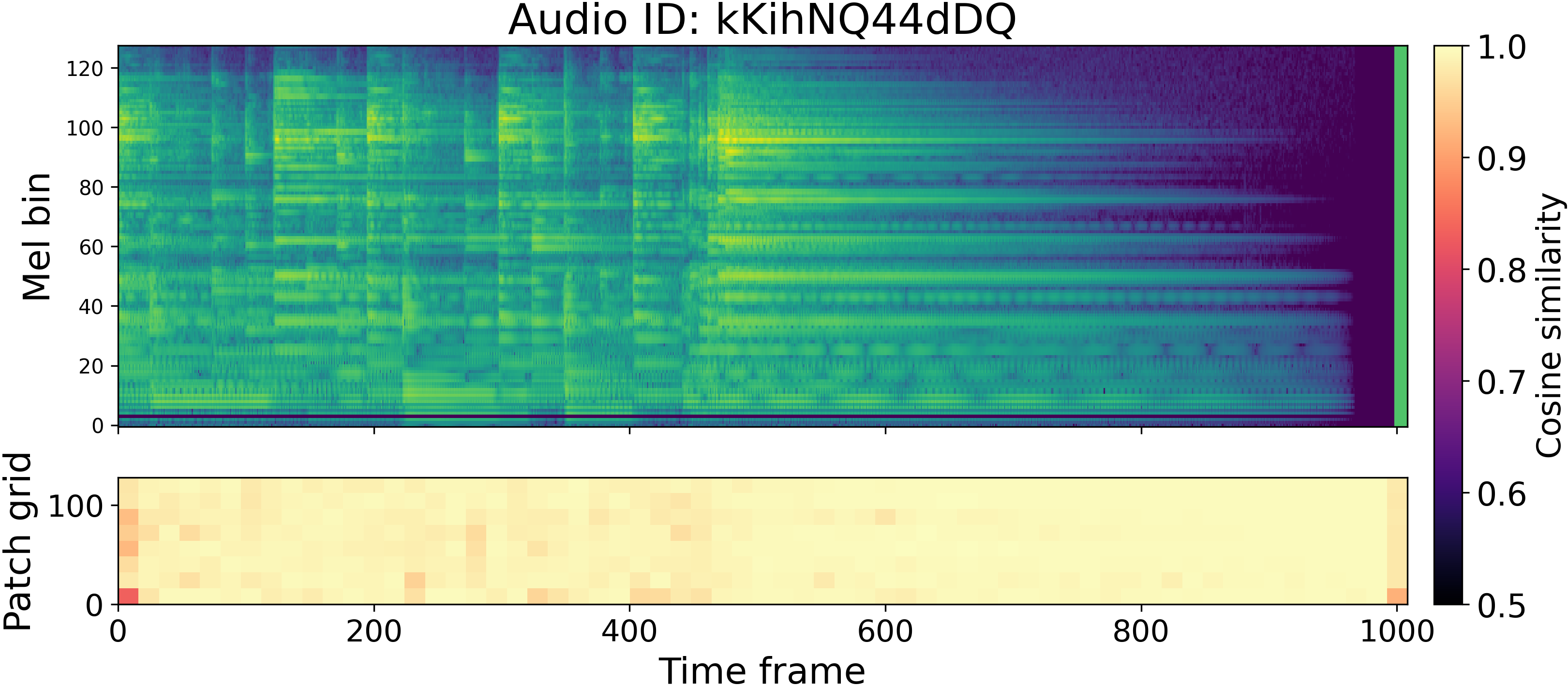}
    \end{subfigure}

    \vspace{0.6em}   % vertical gap between rows

    % --- bottom row: two subfigures ---
    \begin{subfigure}[t]{0.9\textwidth}
        \centering
        \includegraphics[width=\linewidth]{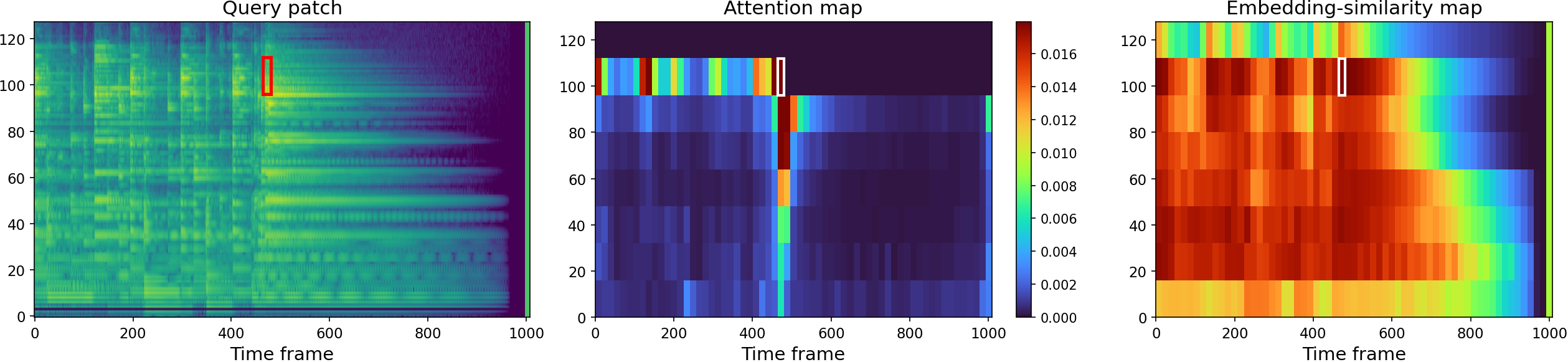}
    \end{subfigure}
    
    \caption{\textbf{Top:} \hlc{green_ablation}{Prediction Quality Analysis.} We report the cosine similarity between the predicted and true patch embeddings, averaged over $500$ audio clips (\textbf{left}) and for two individual clips (\textbf{middle}, \textbf{right}). \textbf{Bottom:}
    \hlc{violetblue_ablation}{Attention/Embedding Analyses.} \textbf{Left:} selected query patch; \textbf{middle:} attention map showing the patches NAPE attends to when predicting the next patch; \textbf{right:} embedding-similarity map between the predicted next-patch embedding $\hat{z}_{t+1}$ and every actual patch embedding in the spectrogram.}
    
    \label{fig:qualitative}
    \vspace{-1.3em}
\end{figure}

\paragraph{\hlc{green_ablation}{Prediction Fidelity.}} We measure the cosine similarity between the predicted embeddings $\hat{z}_{t+1}$ and the true patch embeddings $z_{t+1}$. Figure~\ref{fig:qualitative}~(top left) shows the similarity averaged across $500$ held-out AudioSet clips, and the top middle and right panels report the same measurement for two individual clips: NAPE predicts the next patch embedding accurately almost everywhere on the grid, with similarity close to the ceiling of $1.0$ both on average and per clip. The remaining low-similarity regions have clear structural explanations. The very first patch has the lowest similarity, since no previous context is available for the prediction to condition on. The first mel row is harder on average, since it combines limited past context with low-frequency patches. Finally, the rightmost patches show slightly lower similarity because the last time columns correspond to zero-padded frames appended to reach the target clip length. Away from these boundary regions, NAPE satisfies its pre-training objective on unseen audio.

\paragraph{\hlc{violetblue_ablation}{Attention and Embedding Analysis.}} To understand \emph{how} NAPE arrives at its predictions, we select a query patch (marked in red in Figure~\ref{fig:qualitative}, bottom left) and analyze the attention and embedding structure it induces. The attention map (bottom middle), which conveys the aggregated attention from the query position to every other patch averaged over all layers and heads, reveals a highly structured pattern with two distinct components: NAPE attends strongly to the current time column, integrating the full spectral profile of the current moment, and to same-mel-frequency patches earlier in the clip, tracking the temporal evolution of the frequency it is about to predict. The embedding-similarity map (bottom right), which compares the predicted embedding $\hat{z}_{t+1}$ against every actual patch embedding in the clip, shows that $\hat{z}_{t+1}$ is most similar to patches that share acoustic structure and energy with the query, regions of the spectrogram carrying comparable spectral content, with similarity gradually decaying at the temporal and spectral extremes. This grouping into coherent acoustic components emerges without any explicit labels or region annotations, suggesting that despite being trained only on a local next-patch objective, NAPE develops representations that capture the broader acoustic structure of the clip.

\section{Conclusion}
\label{sec:conclusion}

We presented NAPE, a self-supervised framework for audio
representation learning based on causal next patch embedding
prediction. Departing from the reconstruction- and masking-based
approaches that dominate audio SSL, NAPE relies on a deliberately minimalist recipe: a single causal Transformer encoder, a lightweight predictor head, and a negative cosine similarity loss with stop-gradient on the target branch. Across six audio and speech benchmarks, NAPE achieves strong downstream performance while relying on a substantially simpler pre-training recipe. NAPE also exhibits favorable scaling behavior across three encoder sizes and delivers competitive linear-probing results. A qualitative analysis further shows that the model develops structured attention patterns and organizes its learned embeddings into acoustically coherent regions without any explicit supervision. Together, these results establish autoregressive next-embedding prediction as a simple, scalable, and effective self-supervised objective for audio, and open a direct path to bringing the causal predictive paradigm into audio representation learning.

\subsubsection*{Acknowledgments}
We thank Andrew Rouditchenko (Nvidia) for his insightful and valuable discussions.

\bibliography{iclr2027_conference}
\bibliographystyle{iclr2027_conference}

\clearpage

\appendix
\section{Appendix}
\label{sec:appendix}

\section{Related Work}
\label{sec:relatedwork}

\paragraph{Supervised Audio Pre-training.} Supervised pre-training for audio has been mainly explored in two regimes.
\emph{Out-of-domain} approaches adapt models originally trained on labeled image datasets such as ImageNet~\citep{deng2009imagenet} to spectrogram inputs, typically by modifying the input layer from three RGB channels to a single-channel spectrogram. Early work of this kind used CNN backbones such as EfficientNet~\citep{gong2021psla}; more recent work adopts Transformer-based architectures, notably
AST~\citep{gong2021ast} and HTS-AT~\citep{chen2022hts}, which
have driven substantial gains on audio classification benchmarks. \emph{In-domain} approaches instead pre-train directly on the target modality. CLAP~\citep{elizalde2023clap} adapts the CLIP~\citep{radford2021learning} recipe to audio via contrastive language-audio alignment on supervised text-audio pairs, while Audio-CLIP~\citep{guzhov2022audioclip} and Wav2clip~\citep{wu2022wav2clip} extends the CLIP backbone
with an additional audio encoder trained on
AudioSet~\citep{gemmeke2017audio}. Despite delivering strong
results, these methods rely on large quantities of labeled data, which are expensive and time-consuming to obtain in practice.

\paragraph{Self-Supervised Audio Pre-training.} 
Self-supervised learning has driven substantial progress in audio representation learning, largely by transferring ideas developed in the vision domain. Most recent methods extract log-mel spectrograms as input and follow the masked image modeling paradigm~\citep{he2022masked}: Audio-MAE~\citep{huang2022masked}, MaskSpec~\citep{chong2023masked} and MSM-MAE~\citep{niizumi2022masked} directly applie the MAE reconstruction objective to spectrogram patches; and
BEATs~\citep{chen2022beats} instead trains an iterative acoustic tokenizer to provide discrete semantic prediction targets. Waveform-based approaches such as wav2vec~2.0~\citep{baevski2020wav2vec} and
data2vec~\citep{baevski2022data2vec} bypass spectrogram
preprocessing altogether, and predict latent representations of the raw audio signal. More recent efforts push these directions further by combining ideas from multiple objectives. EAT~\citep{chen2024eat} combines an MAE-style reconstruction with the data2vec latent-target formulation. A-JEPA~\citep{fei2023jepa} adapts the joint-embedding predictive paradigm to audio, using a target encoder updated via EMA and predicting masked patch embeddings. SSLAM~\citep{alex2025sslam} extends EAT with an additional source retention loss trained on artificially mixed audio to improve robustness to polyphonic content. Recently, SPEAR~\citep{yang2025spear} unifies speech and general audio representation learning through the distillation of complementary knowledge from specialized teacher models. Despite their diversity, these methods all rely on bidirectional prediction of masked or corrupted audio, and typically require a reconstruction decoder, a student-teacher setup, or auxiliary regularization losses to achieve competitive performance.

\paragraph{Next-token/embedding Prediction.} Predictive learning has long been a central principle in representation learning across modalities. In language, GPT-style models~\citep{radford2018improving, radford2021learning, achiam2023gpt} established autoregressive next-token prediction as a scalable pre-training objective, with subsequent work confirming that the same paradigm transfers to vision (Image-GPT~\citep{chen2020generative}) and beyond. More recent efforts move away from predicting raw signal
tokens toward predicting embeddings directly. A first line of work adopts an \emph{autoregressive} formulation:
NEPA~\citep{xu2025nepa} introduces next-embedding predictive
autoregression for visual representation learning;
PointNTP~\citep{yao2026rethinking} adapts causal next-token
predictive learning to 3D point clouds; and
LeWorldModel~\citep{maes2026leworldmodel} together with
stable-worldmodel~\citep{maes2026stable} extend the paradigm to next-frame prediction for control tasks. A second line of work adopts a \emph{joint-embedding} formulation: I-JEPA~\citep{assran2023self} and A-JEPA~\citep{fei2023jepa} predict latent representations of masked regions from context using dual encoders and an EMA teacher, while LeJEPA~\citep{balestriero2025lejepa} dispenses with
masking and instead enforces invariance across augmented views with a variance-covariance regularizer. Together, these works suggest that predicting future tokens or embeddings can serve as a unified and scalable pre-training principle across modalities. In audio, however, this paradigm has remained largely unexplored: existing SSL methods rely almost exclusively on bidirectional masked modeling, and it is unclear \emph{a priori} whether causal, per-position prediction is compatible with the non-stationary, temporally
structured nature of audio signals. NAPE is, to the best of our knowledge, the first method to demonstrate that causal next-embedding prediction is a competitive and scalable pre-training objective for audio spectrograms, matching or surpassing more elaborate masked-modeling and joint-embedding
methods with a substantially simpler recipe.

\section{Full Hyperparameter List}
\label{sec:hyperparameters}

\begin{table}[t]
\small
\renewcommand{\tabcolsep}{0.4mm}
\centering
    \caption{\textbf{Hyperparameter list}. When an hyperparameter $h$ varies between the base ($h_b$) and large ($h_l$) model, we include both values like ($h_b$/$h_l$). $^\ast$Following \citep{chen2022beats}, we balance each class to $50$\% of the size of the unknown class for each training epoch. }
%\resizebox{0.999\linewidth}{!}{
\begin{tabular}{lccccccc}
\toprule
\multirow{2}{*}{\textbf{Hyperparameters}} & {\cellcolor{greymiddle}\textbf{Pre-training}} & \multicolumn{6}{c}{\cellcolor{pinksecondbest}\textbf{Fine-tuning}}\\ & \emph{AS-2M} & \emph{AS-2M} & \emph{AS-20K} & \emph{ESC-50} & \emph{KS1} & \emph{KS2} &\emph{ER} \\

\midrule

Optimizer & \multicolumn{7}{c}{AdamW} \\
Opt. Momentum ($\beta_1$, $\beta_2$) & (0.9,0.95) & \multicolumn{6}{c}{(0.9,0.999)} \\
Weight Decay & \multicolumn{7}{c}{0.05} \\
Learning Rate Scheduler & \multicolumn{7}{c}{Cosine Decay} \\
Layer-Wise LR Decay & 1.0 & 0.7/0.9 & 0.8/0.9 & 0.7/0.9 & 0.7/0.8 & 0.7/0.8 & 0.7/0.9 \\
Base Learning Rate & 5e-3 & \multicolumn{6}{c}{1.25e-3}  \\
Epochs & 30/25 & 20/15 & 30/20 & 100 & 50 & 50 & 50 \\
Warm-up Epochs & 3 & 4/3 & 6/5 & 10 & 5 & 5 & 5 \\
Batch Size & 256/128 & \multicolumn{6}{c}{64}  \\
GPUs & 8 & 4 & 4 & 1 & 1 & 1 & 4 \\
Weighted sampling & \ding{55} & \ding{51} & \ding{55} &\ding{55} &\ding{51}$\ast$ &\ding{55} &\ding{55} \\
Multilabel & N/A & \ding{51} & \ding{51} & \ding{55} & \ding{55} & \ding{55} & \ding{55} \\ 
EMA Decay Rate &0.9999 & 0.99995 & 0.999 & \ding{55} & \ding{55} & \ding{55} & \ding{55} \\
Label Smoothing & N/A & 0. & 0. & 0.1 & 0.1 & 0. & 0.1 \\
Roll Augmentation & \ding{55} & \ding{51} & \ding{51} & \ding{51} & \ding{55} & \ding{55} & \ding{51} \\
Drop Path & 0. &  \multicolumn{6}{c}{0.1} \\
SpecAug (time/freq) & N/A & (96,16) & (24,16)/(96,16) & (96,24)/(24,16) & (24,16) & (24,16) & (48,24) \\
Mixup (alpha/prob.) & N/A  & (0.8,1.0) & (0.8,0.8) & (0.8,0.5) & (0.8,0.8) & (0.8,0.8) & (0.8,0.5) \\ 
Cutmix (alpha/prob.) & N/A & (1.0,1.0) & (1.0,0.8) & (1.0,0.5) & (1.0,0.8) & (1.0,0.8) & (1.0,0.5) \\
Noise Augmentation & \ding{55} & \ding{51} & \ding{51} & \ding{51} & \ding{51} & \ding{51} & \ding{51}\\
Loss Function & Neg Cos Sim & BCE & BCE & CE & BCE & BCE & CE \\ 
Dataset Mean for Norm. & -6.84 & -6.84 & -6.84 & -6.84 & -9.11 & -9.16 & -13.74 \\
Dataset Std for Norm. & 5.38 & 5.38 & 5.38 & 5.38 & 4.53 & 4.61 & 3.88\\

\bottomrule

\end{tabular}%}
\label{tab:hyperparameters}
%\vspace{-0.2cm}
\end{table}

\paragraph{Fine-tuning Setting.} We report the full hyperparameter list for fine-tuning NEPA Base and Large in Table \ref{tab:hyperparameters}. Those values refer to the Raster scan variant, with the Diagonal variant having almost the same hyperparameters (the only difference is in the number of epochs needed to converge for \emph{AS-20K}, which can vary of only a few epochs). Regarding NEPA Small, we used the same hyperparameters as NEPA Base, with the only difference being the value of the layer-wise learning rate decay ($LLRD$) for some of the downstream tasks as this hyperparameter depends on the number of layers of the model. Specifically, we set $LLRD = 0.6$ for \emph{AS-2M} and $LLRD = 0.6$ for \emph{KS1}/\emph{KS2}. No other changes have been made.

\paragraph{Linear Probing Setting.} For the linear probing experiments, we made the following changes (all other settings are the same as the fine-tuning hyperparameters): we increase the learning rate to $1e-2$ and we disable all augmentation techniques.

\section{Additional Ablation Studies}
\label{sec:additional_ablations}

In this section, we report additional ablations studies on NAPE. For all experiments we use the raster scan variant. We use the SimSiam predictor for all ablations except for the ablation on the pre-training budget where we use the 2-MLP predictor.

\paragraph{\hlc{orange_ablation}{Freezing the Patch Embedding Layer.}} Table~\ref{tab:patchembfreeze} examines whether freezing the patch embedding layer during fine-tuning affects downstream performance. In~\cite{xu2025nepa}, freezing the embedding layer yields significant improvements. However, in our setting, the two
configurations are essentially indistinguishable on all three tasks tested. We therefore leave the patch embedding layer trainable during fine-tuning, matching the standard practice of prior audio Transformer methods.

\paragraph{\hlc{blue_ablation}{Positional Encoding.}} Table~\ref{tab:pos} compares learned absolute positional encodings with Rotary Position Embedding (RoPE)~\citep{su2024roformer}, applied independently along
the frequency and time axes. RoPE substantially outperforms absolute encodings on both AudioSet benchmarks, with gains of $+1.8$ mAP on AS-2M and $+2.4$ mAP on AS-20K. Beyond the accuracy improvement, RoPE also offers a practical benefit: because it encodes relative positions rather than absolute ones, it transfers naturally to clips of different lengths at inference time, whereas absolute encodings require interpolation. We therefore adopt RoPE as NAPE's default
positional encoding.

\begin{wraptable}{r}{0.395\textwidth}
\small
\vspace{-1.3em}
\renewcommand{\tabcolsep}{0.5mm}
\centering
    \caption{\hlc{green_ablation}{Ablation on the pooling method and attention type at fine-tuning.}}
%\resizebox{0.999\linewidth}{!}{
\begin{tabular}{lcccc}
\toprule
\multicolumn{1}{c}{\textbf{Attention}} & \textbf{Pooling} & \multicolumn{3}{c}{\cellcolor{yellow}\textbf{Task}} \\ \multicolumn{1}{c}{\textbf{Type}} & \textbf{Mode} & \emph{AS-2M} & \emph{AS-20K} & \emph{KS2} \\

\midrule

Bidirec & CLS Tok & \textbf{49.7} & 38.9 & 98.7 \\
Bidirec & Last Tok & 49.6 & 38.7 & \textbf{98.8} \\
\textbf{Bidirec} & \textbf{Avg Pool} & 49.6 & \textbf{39.1} & \textbf{98.8} \\
\hline \addlinespace[2pt] 
Causal & Last Tok & 49.4 & 38.9 & \textbf{98.8} \\

\bottomrule
\end{tabular}%}
\label{tab:causalbid_table}
%\vspace{-0.2cm}
\end{wraptable}

\paragraph{\hlc{green_ablation}{Attention Type and Pooling.}}
Table~\ref{tab:causalbid_table} jointly ablates the attention mask used during fine-tuning and the pooling strategy for producing the clip-level representation. Under causal attention, the natural pooling choice is the last token: only the final position has access to the full preceding context, and the CLS token functions more as a BOS-like anchor at the start of the sequence than as a readout summary of the clip (this is in line with \citep{xu2025nepa}). Under bidirectional attention, however,
every position sees the entire sequence, so CLS, last-token, and mean pooling all in principle have access to the same information. This is reflected in the numbers: the three bidirectional variants are essentially on par on the three tested tasks. Keeping causal attention with last-token pooling remains competitive but slightly worse than its bidirectional counterpart, with a $0.2$ mAP drop on both AS-2M and AS-20K. We therefore use bidirectional attention with mean pooling as NAPE's default fine-tuning configuration.

\begin{table}[t]
\renewcommand{\tabcolsep}{0.5mm}
\centering
\begin{minipage}[t]{0.50\textwidth}
    \centering
    \caption{\hlc{orange_ablation}{Ablation on freezing the emb. layer.}}
    \label{tab:patchembfreeze}
    \begin{tabular}{cccc}
    \toprule
    \textbf{Freeze emb} & \emph{AS-2M} & \emph{AS-20K} & \emph{KS2}\\
    \midrule
    \ding{51} & 49.61 & 39.14 & 98.73 \\
    \ding{56} & 49.58 & 39.08 & 98.80 \\
    \bottomrule
    \end{tabular}
\end{minipage}
\hfill
\begin{minipage}[t]{0.46\textwidth}
    \centering
    \caption{\hlc{blue_ablation}{Ablation on positional encoding.}}
    \label{tab:pos}
    \begin{tabular}{lcc}
    \toprule
    \multicolumn{1}{c}{\textbf{Positional Enc.}} & \emph{AS-2M} & \emph{AS-20K}\\
    \midrule
    Absolute & 47.8 & 36.7 \\
    \textbf{RoPE} & \textbf{49.6} & \textbf{39.1} \\
    \bottomrule
    \end{tabular}
\end{minipage}
\end{table}

\begin{wraptable}{r}{0.445\textwidth}
\small
\vspace{-1.3em}
\renewcommand{\tabcolsep}{0.5mm}
\centering
    \caption{\hlc{pink_ablation}{Ablation on normalization layer style.}}
%\resizebox{0.999\linewidth}{!}{
\begin{tabular}{lcccccc}
\toprule
\multicolumn{1}{c}{\textbf{Norm}} & \multicolumn{3}{c}{\textbf{Audio Tasks}}  & \multicolumn{3}{c}{\textbf{Speech Tasks}}\\ \multicolumn{1}{c}{\textbf{Type}} & \emph{AS-2M} & \emph{AS-20K} & \emph{ESC-50} & \emph{KS1} & \emph{KS2} &\emph{ER} \\ 

\midrule

\textbf{LN} & 49.6 & 39.1 & 94.2 & 97.9 & 98.8 & 64.9 \\
RMS & 49.5 & 38.8 & 94.4 & 97.8 & 98.8 & 65.2 \\

\bottomrule

\end{tabular}%}
\label{tab:norm}
%\vspace{-0.2cm}
\end{wraptable}

\paragraph{\hlc{pink_ablation}{Normalization Layer.}} Table~\ref{tab:norm} compares LayerNorm~\citep{ba2016layer} (LN) and RMSNorm~\citep{zhang2019root} in the encoder. RMSNorm drops the mean-centering step of LayerNorm and normalizes each feature vector by its root-mean-square, which has become common in recent language and vision Transformers. In our setting, the two variants are effectively on par, with small variations across the six downstream tasks. For this reason, we retain LayerNorm as NAPE's default for consistency with prior audio Transformer works.

\begin{wrapfigure}{r}{0.5\textwidth}
    \vspace{-0.3cm}
    \centering
    \includegraphics[width=\linewidth]{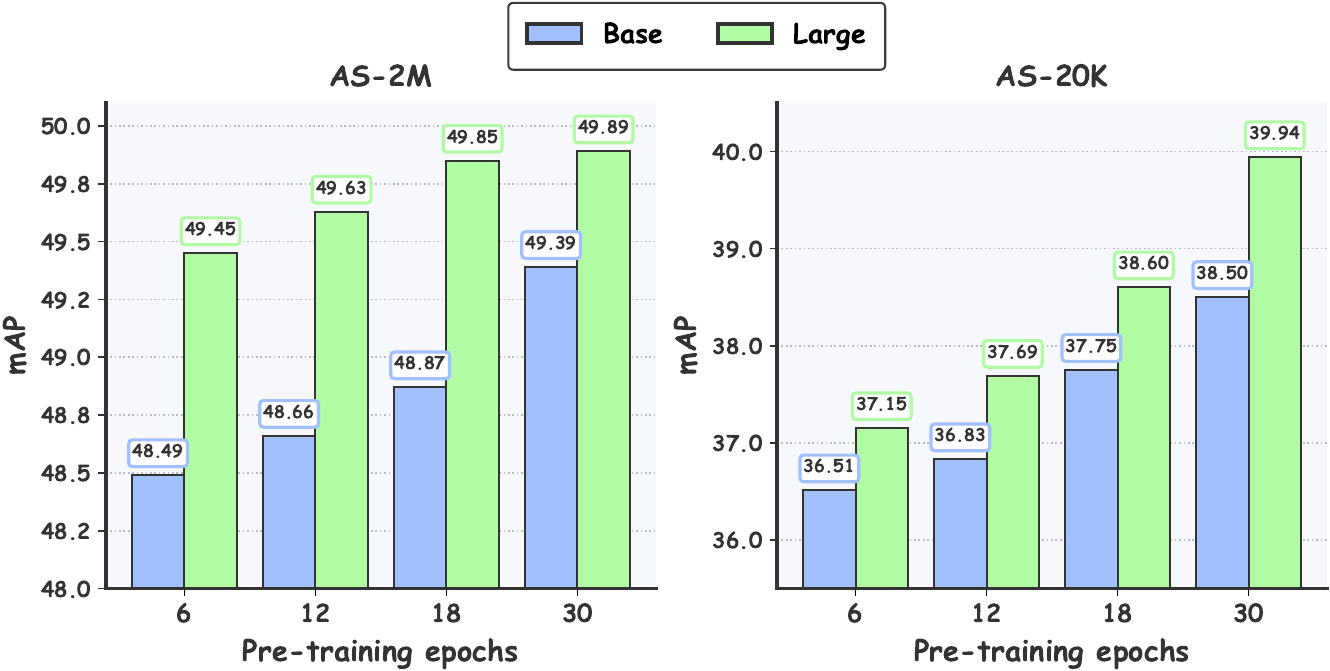}
    \caption{\hlc{yellow_ablation}{Pre-training budget ablation}. AS-2M (left) and AS-20K (right) mAP trend as a function of the number of epochs the model is trained on.}
    \label{fig:epochs_trend}
    \vspace{-0.3cm}
\end{wrapfigure}
\paragraph{\hlc{yellow_ablation}{Pre-training Budget.}} Figure~\ref{fig:epochs_trend} reports downstream mAP on AS-2M and AS-20K when fine-tuning from NAPE-Base and NAPE-Large checkpoints obtained after $6$, $12$, $18$, and $30$ epochs of pre-training. Performance improves monotonically with the pre-training budget for both scales and both benchmarks: on AS-2M, NAPE-Large improves from $49.45$ mAP after $6$ epochs to $49.89$ after $30$; on AS-20K, from $37.15$ to $39.94$. Notably, NAPE already delivers strong results after only a few epochs of pre-training. After just $6$ epochs, NAPE-Large reaches $49.45$ mAP on AS-2M — already matching or exceeding most published baselines (cf.\ Table~\ref{tab:main_table}) — making NAPE an appealing choice under tight compute budgets. Longer pre-training continues to yield returns at $30$ epoch, and NAPE-Large consistently benefits more than NAPE-Base from a larger budget, consistent with the scaling behavior discussed in Section~\ref{sec:scaling}.

\begin{wraptable}{r}{0.43\textwidth}
\small
\vspace{-1.5em}
\renewcommand{\tabcolsep}{0.5mm}
\centering
    \caption{\hlc{grey_ablation}{Ablation on random masking applied to the input embeddings.}}
%\resizebox{0.999\linewidth}{!}{
\begin{tabular}{ccccccc}
\toprule
\multicolumn{1}{c}{\textbf{Mask}} & \multicolumn{3}{c}{\textbf{Audio Tasks}}  & \multicolumn{3}{c}{\textbf{Speech Tasks}}\\ \multicolumn{1}{c}{\textbf{Ratio}} & \emph{AS-2M} & \emph{AS-20K} & \emph{ESC-50} & \emph{KS1} & \emph{KS2} &\emph{ER} \\ 

\midrule

\textbf{0} & \textbf{47.6} & \textbf{36.2} & \textbf{92.9} & 97.4 & 98.3 & \textbf{64.6} \\
20 & 47.3 & \textbf{36.2} & 92.2 & \textbf{97.6} & \textbf{98.5} & 63.9 \\
50 & 47.2 & 35.2 & 91.7 & 97.5 & 98.3 & 63.9 \\

\bottomrule

\end{tabular}%}
\label{tab:mask}
\vspace{-1em}
\end{wraptable}
\paragraph{\hlc{grey_ablation}{Random Masking.}} Table~\ref{tab:mask} examines whether adding random masking to the input embeddings during pre-training,
in the style of masked modeling methods, helps NAPE. We consider three masking ratios: no masking ($0\%$, the default), $20\%$, and $50\%$. We use NAPE small with raster scan order.  Across all six benchmarks, masking either has no
effect or slightly degrades downstream performance, with the
largest drop on AS-20K ($-1.0$ mAP at $50\%$) and ESC-50
($-1.2$ points at $50\%$). This differs from methods such as
Audio-MAE~\citep{huang2022masked} or BEATs~\citep{chen2022beats}, where high masking ratios are central to the pre-training signal: the model is asked to reconstruct or predict the masked content from the visible context. In NAPE, the causal attention mask already prevents each position from accessing its target, so additional input masking removes useful context without changing the difficulty of the prediction task. We therefore leave the
input unmasked in NAPE's default configuration.

\section{NAPE vs JEPA vs LeWorldModel Formulation}
\label{sec:napevsjepa}

\subsection{NAPE vs JEPA-style Methods} NAPE and JEPA methods~\cite{assran2023self, fei2023jepa} learn representations by predicting one set of patch embeddings from
another, therefore it is worth clarifying how the two paradigms differ structurally. Figure~\ref{fig:nape_jepa_leworldmodel} (left and right) illustrates the two frameworks.

JEPA-style methods use \emph{two} encoders: a context encoder $h$ that processes the visible portion of the input $x$, and a target encoder that produces the prediction targets from the complementary region $y$. The target encoder is not trained by gradient descent; instead, its weights are updated as an EMA of the context encoder's weights, so that its output remains a moving target that the predictor tries to match. The predictor itself is a Transformer that receives both the context representation and an additional variable
$c$, which encodes the target positions to be predicted, and
produces the predicted target embeddings $\hat{z}_y$ in parallel. The context encoder operates with bidirectional attention over the visible patches. To prevent representation collapse under this regime, JEPA methods typically require additional machinery beyond stop-gradient: an EMA teacher (I-JEPA~\citep{assran2023self}, A-JEPA~\citep{fei2023jepa}), auxiliary variance-covariance regularizers such as VICReg~\citep{bardes2021vicreg}, or masking strategies carefully tuned so that the two views do not degenerate
into trivial solutions.

NAPE, in contrast, uses a \emph{single} encoder ($h$) and takes its prediction targets directly from the shared patch embedding layer $f$. Because $f$ is a shallow, non-recurrent module, its outputs provide a stable target signal that does not need to be maintained by a separate EMA-tracked encoder. NAPE also features a lightweight MLP head ($g$) that operates on a single position at a time, rather than a Transformer that jointly reasons about context and target positions: a structurally simpler design that reflects NAPE's per-position causal prediction task. There is no masking: the model processes the full input as a causal sequence, and predictions are made one patch at a time under a causal attention mask. Collapse prevention relies solely on the three key components discussed in Section~\ref{sec:nape-prediction}: causality, prediction shift, and stop-gradient, with no auxiliary regularizers or student-teacher setups.

\subsection{NAPE vs LeWorldModel Methods}
\begin{figure}[t]
    \centering
    \includegraphics[width=1\textwidth]{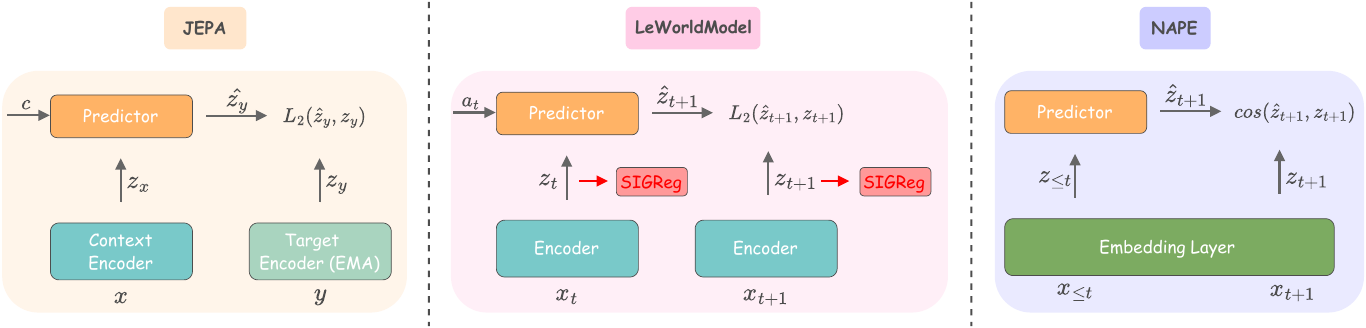}
    \caption{Comparison between JEPA, LeWorldModel, and NAPE architectures.}
    \label{fig:nape_jepa_leworldmodel}
    %\vspace{-0.3cm}
\end{figure}

A closely related recent method is
LeWorldModel~\citep{maes2026leworldmodel}, a JEPA designed for world modeling that also autoregressively predicts next-frame embeddings from raw pixels (Figure~\ref{fig:nape_jepa_leworldmodel}, middle). Like NAPE, LeWM aims to simplify the JEPA recipe and dispenses with
EMA teachers and pretrained encoders, training its encoder and
predictor jointly end-to-end. Unlike NAPE, LeWM relies on the SIGReg regularizer~\citep{balestriero2025lejepa}, an auxiliary loss that projects the learned embeddings onto random directions and enforces Gaussian-distributed marginals, to prevent collapse, whereas NAPE requires only stop-gradient on the target branch. LeWM uses MSE as
its prediction loss (kept stable by SIGReg), whereas NAPE uses
cosine similarity combined with stop-gradient, which together avoid the shrink-to-zero collapse mode observed with $\ell_1$ and $\ell_2$ under a magnitude-sensitive loss (Section~\ref{sec:ablations}). Finally, the two methods target different problems: LeWM predicts across time steps of an action-conditioned trajectory for planning in latent space, while NAPE predicts across patches within a single input for representation learning applicable to classification
tasks. NAPE therefore constitutes a distinct point in the JEPA
design space: one that shares LeWM's end-to-end simplicity but
arrives at collapse prevention through a fundamentally different mechanism.

\section{Training Loss Visualizations}
\label{sec:loss_visualization}

\begin{figure}[t]
    \centering
    % --- top row: four subfigures ---
    \begin{subfigure}[t]{0.325\textwidth}
        \centering
        \includegraphics[width=\linewidth]{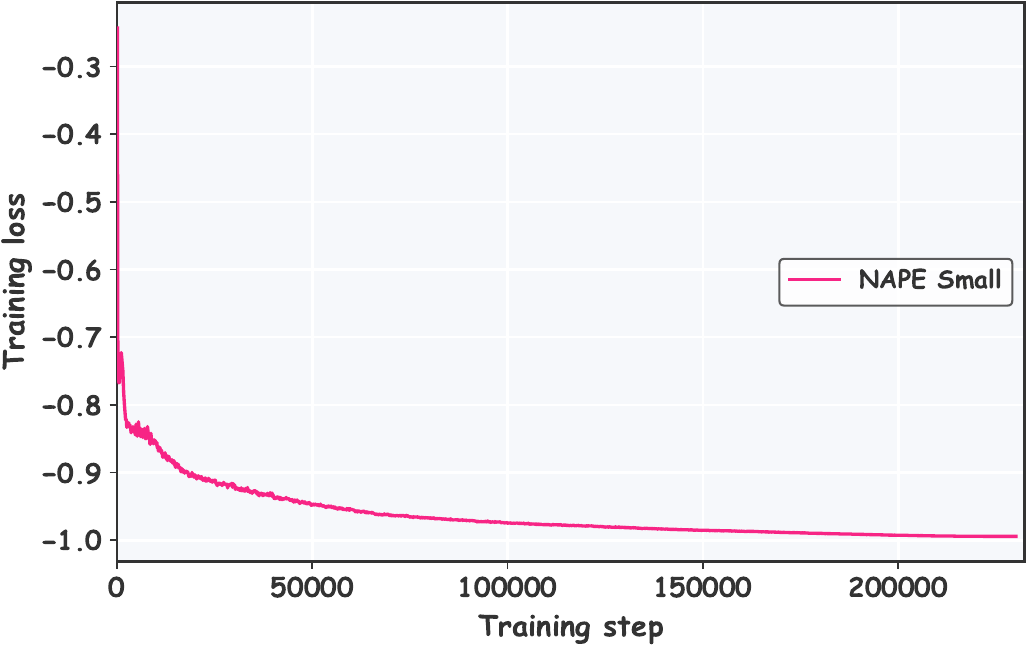}
        %\caption{}\label{fig:a}
    \end{subfigure}
    \hfill
    \begin{subfigure}[t]{0.325\textwidth}
        \centering
        \includegraphics[width=\linewidth]{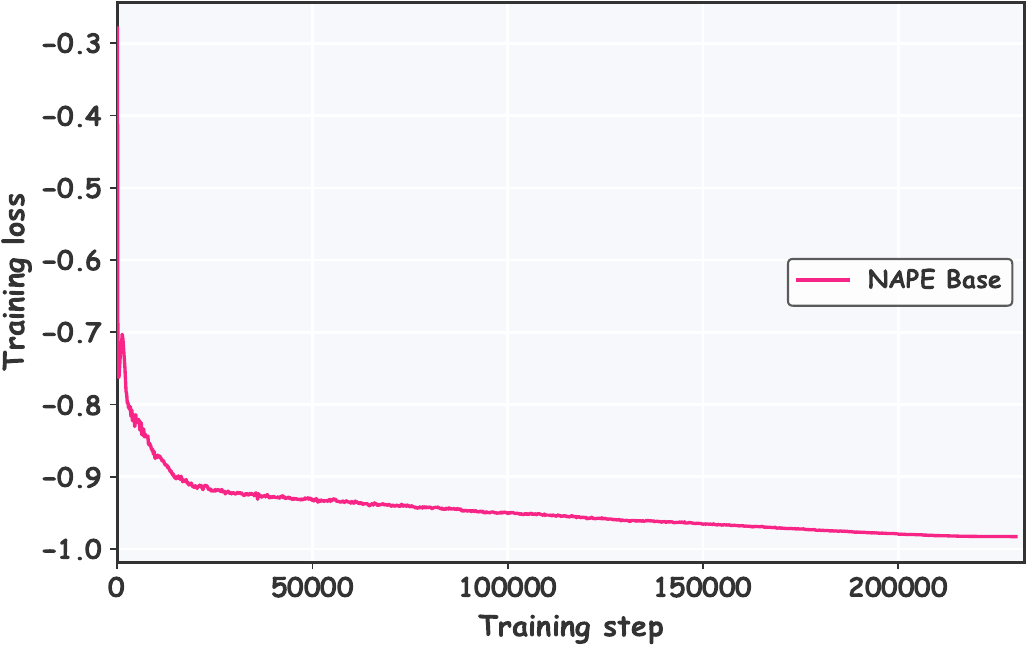}
        %\caption{}\label{fig:b}
    \end{subfigure}
    \hfill
    \begin{subfigure}[t]{0.325\textwidth}
        \centering
        \includegraphics[width=\linewidth]{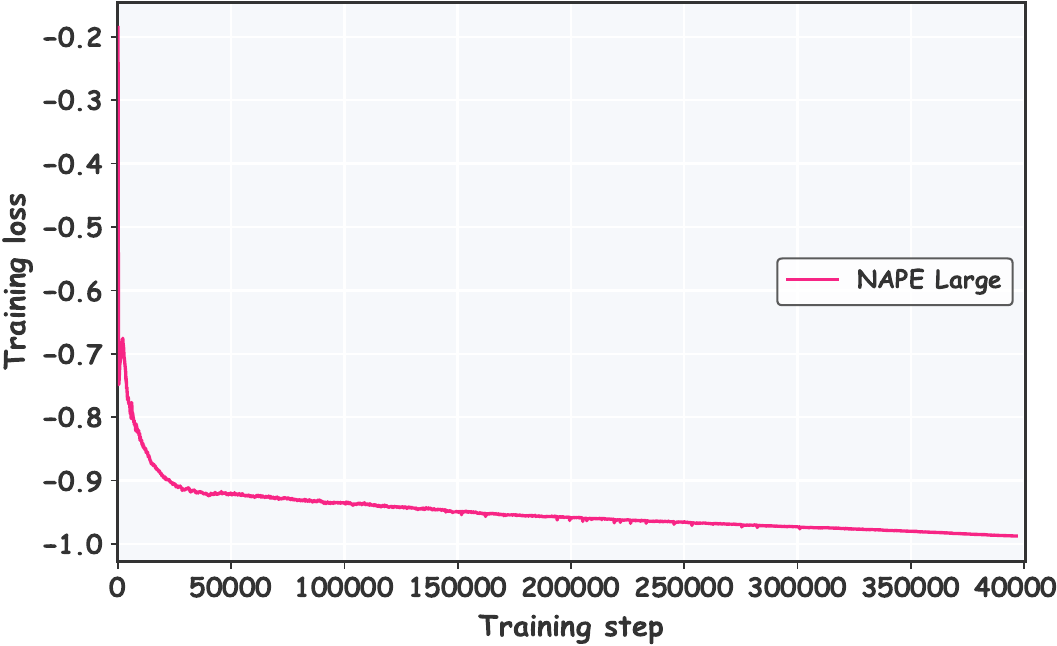}
        %\caption{}\label{fig:b}
    \end{subfigure}

    \vspace{0.6em}   % vertical gap between rows

    % --- bottom row: two subfigures ---
    \begin{subfigure}[t]{0.325\textwidth}
        \centering
        \includegraphics[width=\linewidth]{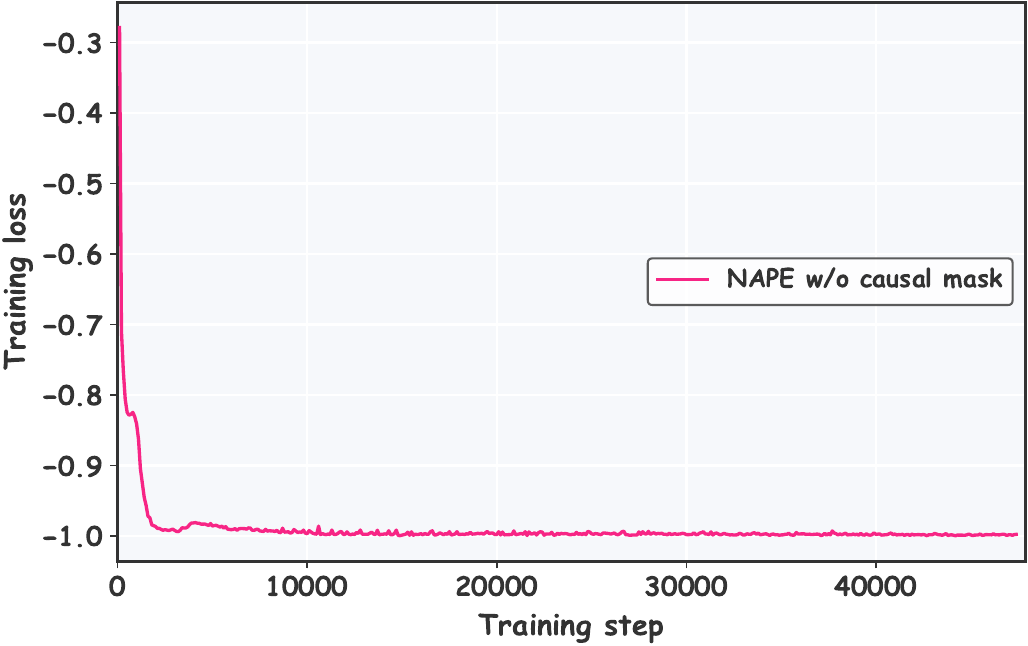}
        %\caption{}\label{fig:a}
    \end{subfigure}
    \hfill
    \begin{subfigure}[t]{0.325\textwidth}
        \centering
        \includegraphics[width=\linewidth]{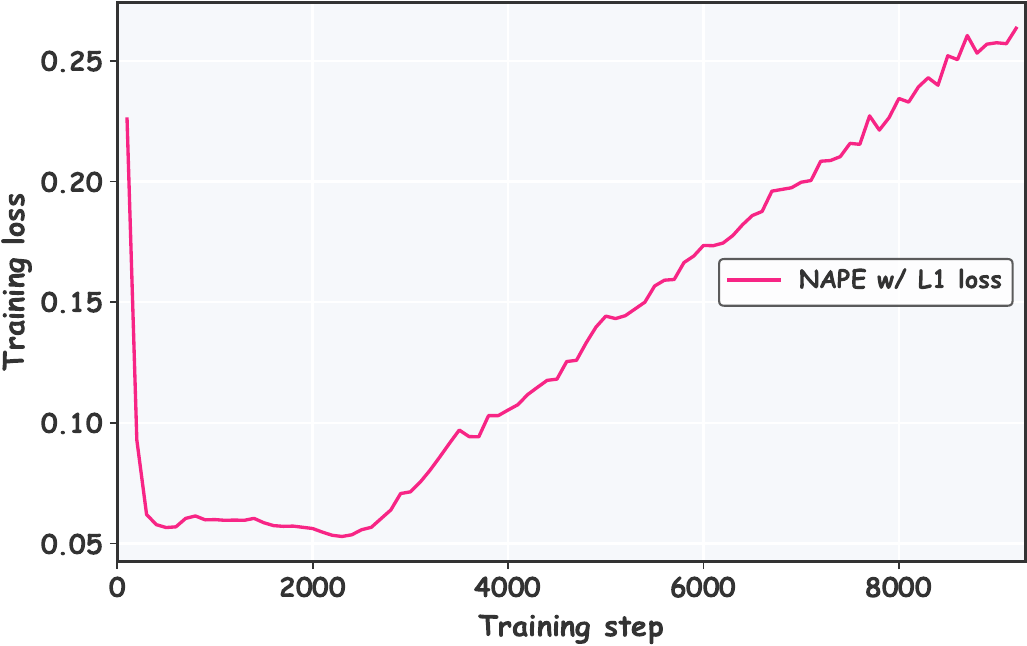}
        %\caption{}\label{fig:b}
    \end{subfigure}
    \hfill
    \begin{subfigure}[t]{0.325\textwidth}
        \centering
        \includegraphics[width=\linewidth]{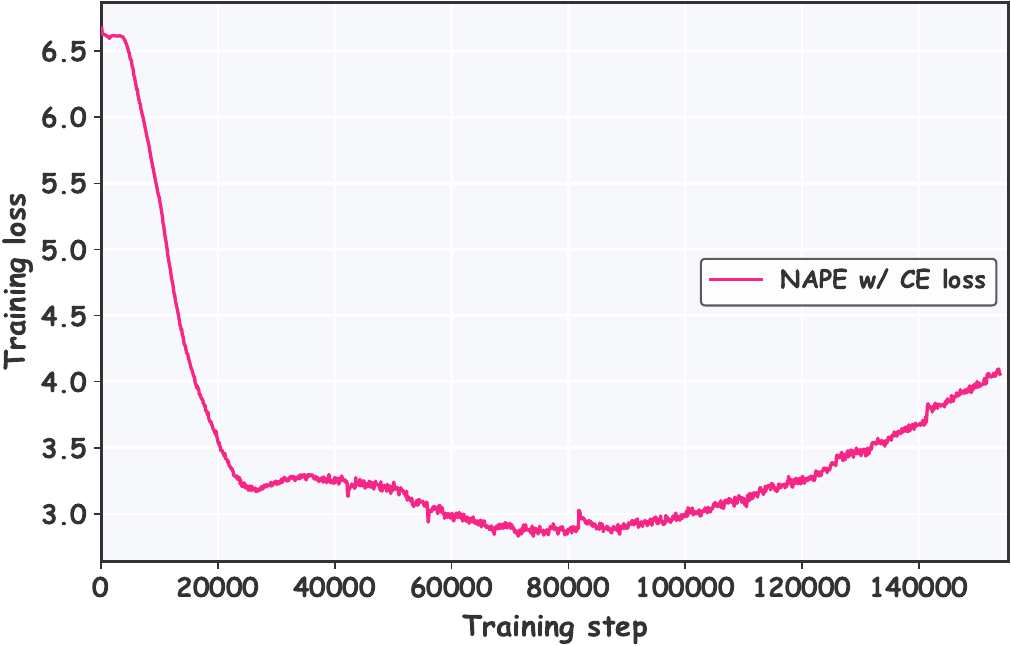}
        %\caption{}\label{fig:b}
    \end{subfigure}

    \caption{Pre-training loss curves for multiple NEPA models across scales and configurations.}
    \label{fig:loss_curves}
\end{figure}

To better understand NAPE's training dynamics, in Figure~\ref{fig:loss_curves} we report the pre-training loss curves of six NAPE runs across multiple scales and configurations. The top row reports runs of the default NAPE configuration (raster scanning, SimSiam predictor, cosine similarity, all three key components enabled) at the three model scales — \textbf{Small} (top left), \textbf{Base} (top middle), and \textbf{Large} (top right). All three curves follow the same qualitative trajectory: a rapid descent from around $-0.3$ to below $-0.9$ within the first $\sim 20{,}000$ steps, followed by a smooth and steady decrease that asymptotes just below $-0.98$, close to the theoretical minimum of $-1$. The Large model runs for roughly twice as many steps as Small and Base, its batch size is halved to fit into memory, but reaches a comparable final loss along a trajectory of the same shape, indicating that the NAPE
objective is well-conditioned across encoder capacities.

The bottom row reports three ablation runs at the Base scale.
\textbf{Without the causal mask} (bottom left), the loss collapses almost immediately, saturating near $-1.0$ within $\sim 2{,}000$ steps and remaining there. The model has found the trivial identity mapping enabled by bidirectional attention: the objective is satisfied without learning any structure that transfers downstream, as reflected in the sharp mAP drop reported in Table~\ref{tab:maincomponents}.

The configuration with \textbf{L1 loss} (bottom middle) diverges: the loss drops rapidly to $\sim 0.05$ within a few hundred steps, plateaus briefly, and then climbs steadily. The initial descent reflects the model finding low-norm solutions in which both predicted and target embeddings shrink toward zero, since L1 is magnitude-sensitive and trivially reduced by scaling the outputs; once the encoder approaches this degenerate regime, the objective becomes ill-conditioned and the loss reverses. The \textbf{cross-entropy loss} (bottom right), in contrast, converges to a reasonable minimum around step $\sim 70{,}000$ before gradually overfitting. Unlike L1, CE does not collapse: this suggests that collapse prevention is not exclusive to cosine similarity. In our setting, however, CE still yields worse downstream performance than cosine (Table~\ref{tab:loss}), and its tendency to overfit after the initial descent makes it a less robust choice for NAPE pre-training.

\section{Additional Qualitative Results}
\label{sec:additional_qualitative}

\begin{figure}[t]
    \centering
    \begin{subfigure}[t]{0.9\textwidth}
        \centering
        \includegraphics[width=\linewidth]{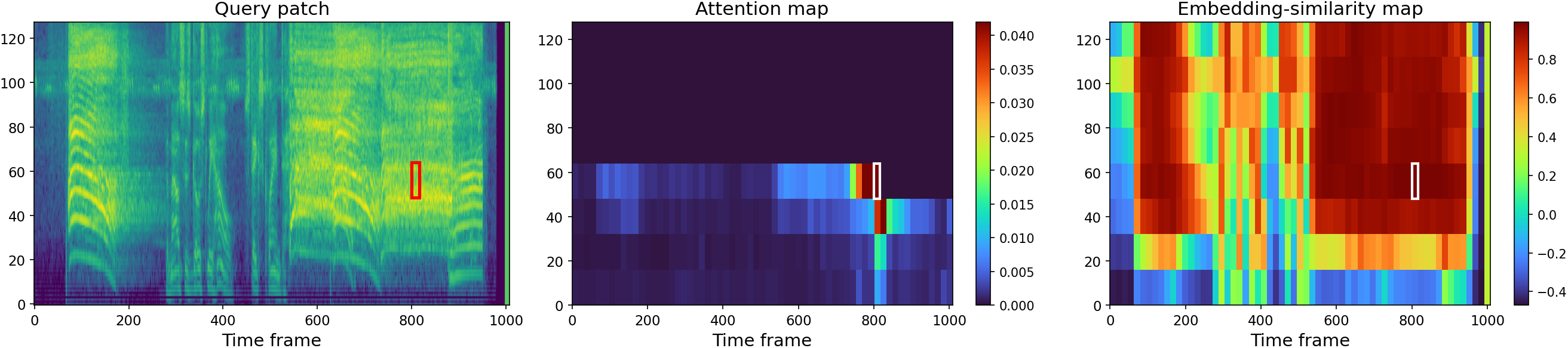}
    \end{subfigure}
    
    \vspace{0.6em}   % vertical gap between rows

    \begin{subfigure}[t]{0.9\textwidth}
        \centering
        \includegraphics[width=\linewidth]{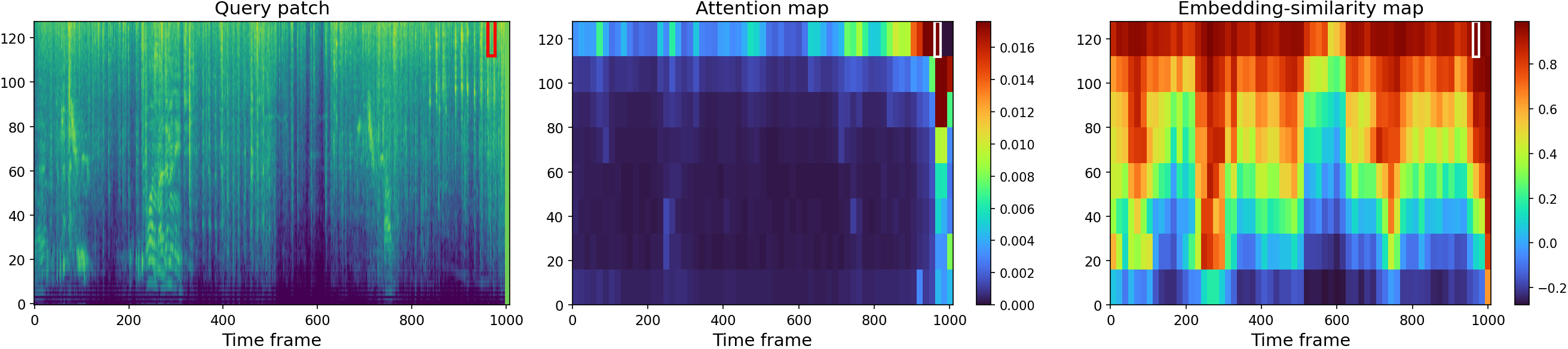}
    \end{subfigure}
    
    \vspace{0.6em}   % vertical gap between rows
    
    % --- bottom row: two subfigures ---
    \begin{subfigure}[t]{0.9\textwidth}
        \centering
        \includegraphics[width=\linewidth]{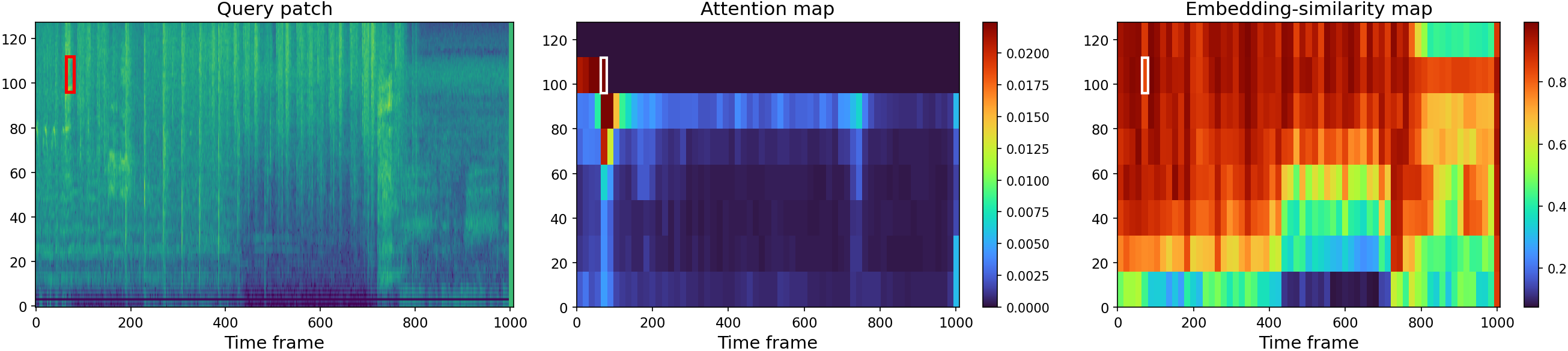}
    \end{subfigure}
    
    \caption{Additional qualitative results for NAPE-L with raster order.}
    
    \label{fig:qualitative_2_raster}
\end{figure}

\begin{figure}[t]
    \centering
    
    \begin{subfigure}[t]{0.9\textwidth}
        \centering
        \includegraphics[width=\linewidth]{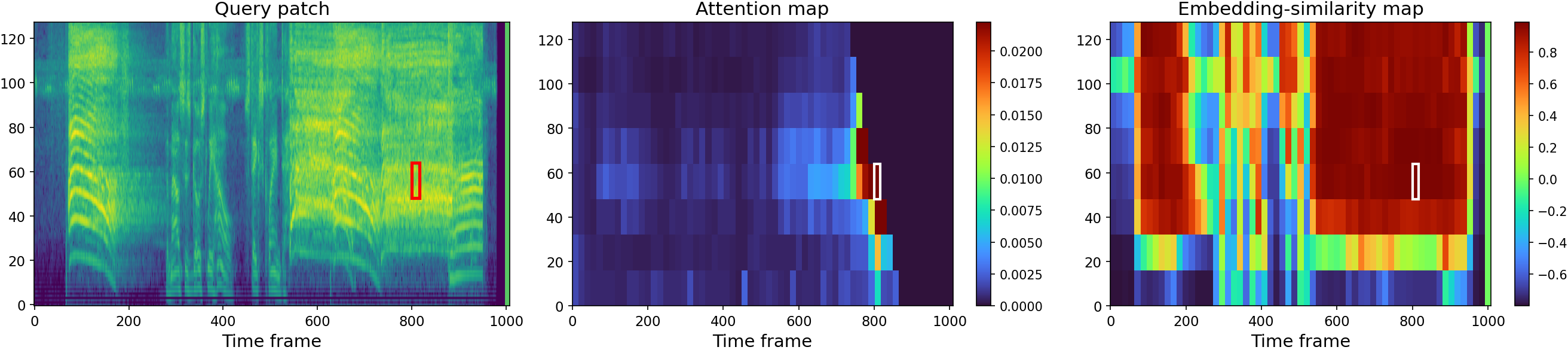}
    \end{subfigure}
    
    \vspace{0.6em}   % vertical gap between rows

    \begin{subfigure}[t]{0.9\textwidth}
        \centering
        \includegraphics[width=\linewidth]{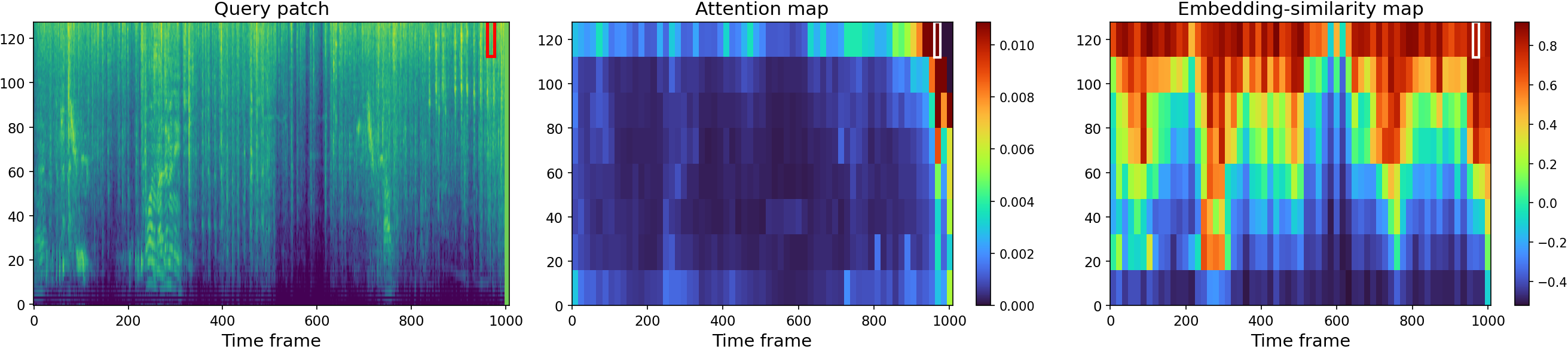}
    \end{subfigure}
    
    \vspace{0.6em}   % vertical gap between rows
    
    % --- bottom row: two subfigures ---
    \begin{subfigure}[t]{0.9\textwidth}
        \centering
        \includegraphics[width=\linewidth]{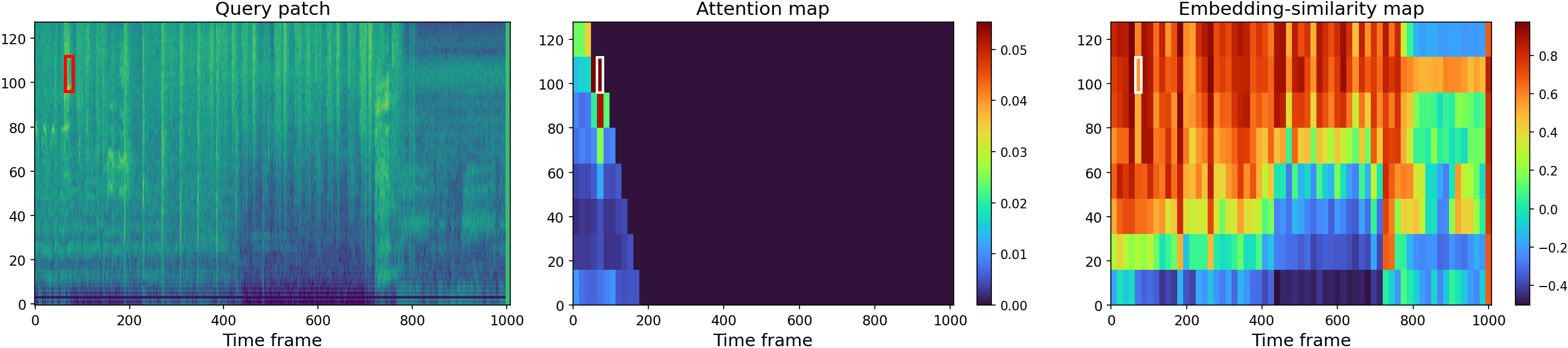}
    \end{subfigure}
    
    \caption{Additional qualitative results for NAPE-L with diagonal order.}
    
    \label{fig:qualitative_2_diagonal}
\end{figure}

We include additional qualitative results for NAPE-L with raster order in Figure~\ref{fig:qualitative_2_raster}) and NAPE-L with diagonal order in Figure~\ref{fig:qualitative_2_diagonal}).

\makeatletter\setlength{\@fptop}{0pt}\makeatother

\end{document}